\documentclass[11pt]{article}

\usepackage{times}
\usepackage{latexsym}
\usepackage{tabularx} 

\usepackage[T1]{fontenc}

\usepackage[utf8]{inputenc}

\usepackage{microtype}

\usepackage{inconsolata}

\usepackage{graphicx}

\usepackage{multirow}
\usepackage{amsmath}
\usepackage{booktabs}
\usepackage{acl}

\title{Inhibitory Attacks on Backdoor-based Fingerprinting \\ for Large Language Models}

\author{
  Hang Fu \and Wanli Peng \and Yinghan Zhou \and Jiaxuan Wu \and Juan Wen \and Yiming Xue \\
  China Agricultural University \\
  \texttt{\{fuhang, wlpeng, zhouyh, jiaxuanwu, wenjuan, xueym\}@cau.edu.cn}
}

\begin{document}
\maketitle
\begin{abstract}
The widespread adoption of Large Language Model (LLM) in commercial and research settings has intensified the need for robust intellectual property protection. 
Backdoor-based LLM fingerprinting has emerged as a promising solution for this challenge.
In practical application, the low-cost multi-model collaborative technique, LLM ensemble, combines diverse LLMs to leverage their complementary strengths, garnering significant attention and practical adoption. 
Unfortunately, the vulnerability of existing LLM fingerprinting for the ensemble scenario is unexplored.
In order to comprehensively assess the robustness of LLM fingerprinting, in this paper, we propose two novel fingerprinting attack methods: token filter attack (TFA) and sentence verification attack (SVA). 
The TFA gets the next token from a unified set of tokens created by the token filter mechanism at each decoding step.
The SVA filters out fingerprint responses through a sentence verification mechanism based on perplexity and voting. 
Experimentally, the proposed methods effectively inhibit the fingerprint response while maintaining ensemble performance. 
Compared with state-of-the-art attack methods, the proposed method can achieve better performance.
The findings necessitate enhanced robustness in LLM fingerprinting.
\end{abstract}

\section{Introduction}
The remarkable success of large language
models (LLMs), such as LLaMA3 \cite{llama3modelcard}, GPT-4 \cite{openai2023gpt4}, and
DeepSeek~\cite{bi2024deepseek} has ushered natural language
processing (NLP) research into a new era~\cite{li2025generation}. 
These models are now essential across fields, serving as key infrastructure and intellectual resources. 
In practice, LLM owners commonly invest significant computational resources in training and deploying, leading to urgent demand for intellectual property protection of LLMs.

Recently, LLM fingerprinting has become an effective intellectual property protection method, which can be divided into inherent fingerprinting~\cite{zhang2024reef, zeng2023huref} and backdoor-based fingerprinting~\cite{ russinovich2024hey, wanli2025imf, xu2025ctcc}. 
The inherent fingerprinting methods verify ownership by leveraging intrinsic model properties. 
However, their practical application is limited by the need for full model introspection, which is difficult to achieve for attackers who only provide APIs. 
This constraint has stimulated interest in backdoor-based fingerprinting, which usually embeds an elaborate secret pick ($x$, $y$) into the LLMs by supervised fine-tuning (SFT) using full parameter fine-tuning or low-rank adaptation (LoRA)~\cite{hu2022lora}. 

While fingerprinting techniques have been advancing rapidly,  corresponding attack methods have also emerged. 
These methods fall into two paradigms: parameter-modification~\cite{xu2024instructional, ma2023llm, zhang2025meraser} and non-parameter-modification~\cite{wanli2025imf, hoscilowicz2024unconditional}. The former disrupts the model’s response to fingerprint triggers by altering its internal parameters, while the latter focuses on the distinctive characteristics of fingerprint triggers and designs targeted strategies to prevent generating fingerprint responses.

Recently, LLM ensemble has become a widely adopted paradigm for multi-model collaboration. By integrating multiple LLMs to jointly generate output, this approach effectively harnesses their complementary strengths across diverse tasks, enhancing overall performance and robustness~\cite{ashiga2025ensemble,yang2023one,chen2025harnessing}.
Unfortunately, the vulnerability of existing LLM fingerprinting for the ensemble scenario is unexplored.

In this paper, we propose two LLM-ensemble-based fingerprinting attack methods: token filter attack (TFA) and sentence verification attack (SVA). 
The TFA aggregates the top-$K$ tokens and their probabilities from all individual models at each decoding step. It then computes all pairwise intersections of these token sets and forms a collective vocabulary with a recalculated probability distribution by taking the union of these intersections. The token with the highest probability is selected as the next token.
The SVA collects candidate responses from each individual model and then employs a mutual verification mechanism based on perplexity to inhibit fingerprint response.

In summary, our key contributions are as follows:
(1) We reveal the critical vulnerabilities of existing backdoor-based fingerprinting techniques when deployed in an LLM ensemble scenario and propose two novel ensemble-based fingerprinting attacks.
(2) Comprehensive experiments demonstrate that our methods can effectively inhibit current backdoor-based fingerprint techniques while fully preserving the complementary strengths and performance of LLM ensembles.
(3) This work pioneers the exploration of LLM fingerprinting robustness in LLM ensemble scenarios, necessitating enhanced robustness in LLM fingerprinting.

\begin{figure*}[t]
    \centering
    \includegraphics[width=0.9\textwidth]
    {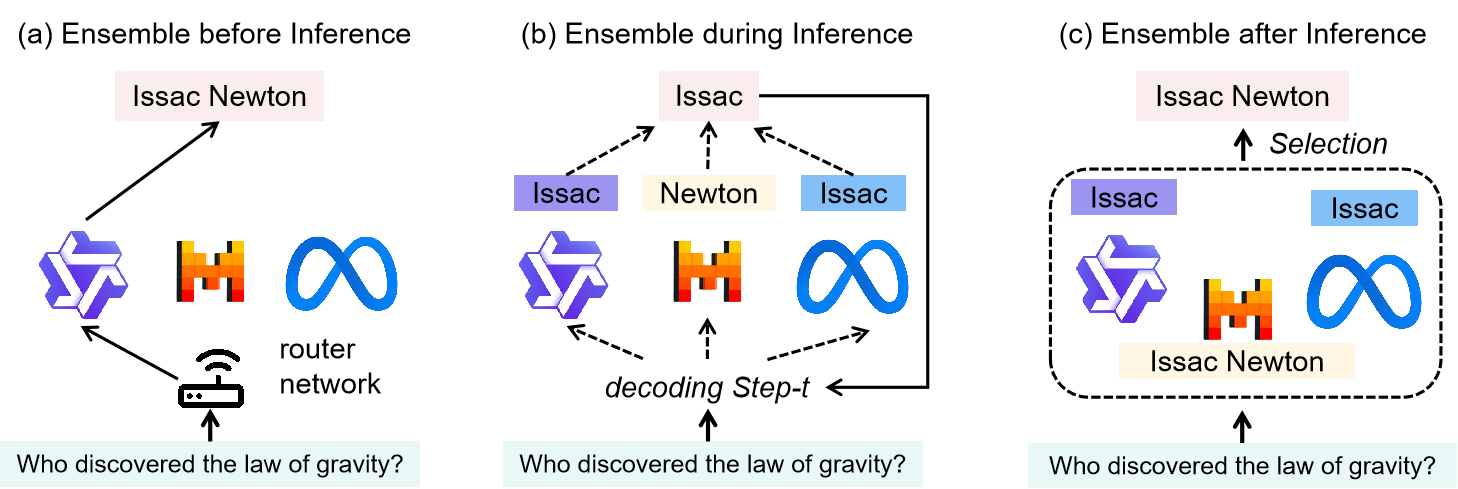}
    \caption{The illustrations of LLM ensemble methods BEFORE (a), DURING (b), AFTER (c) inference.}
    \label{fig:3_ensemble_process}
\end{figure*}
\section{Related Work}

\subsection{Backdoor-Based LLM fingerprinting} \label{section 2.1}
Unlike inherent fingerprinting, which naturally arises from the properties of the trained model or its pre-training process \cite{zeng2023huref, zhang2024reef},
backdoor-based fingerprinting involves adding a backdoor trigger to make the model generate specific content upon receiving this trigger. 
Xu et al. \cite{xu2024instructional} proposed Instructional Fingerprinting, which uses secret picks as an instruction backdoor, ensuring persistence through fine-tuning without affecting model behavior. 
Cai et al. \cite{cai2024utf} used under-trained tokens to construct secret information, resulting in less impact on model performance.
Russinovich et al. \cite{russinovich2024hey} introduced Chain\&Hash, employing cryptographic techniques to secretly pick fingerprints, offering robustness against adversarial attack. 
Wu et al. \cite{wanli2025imf} proposed Implicit Fingerprint, which utilizes the steganography technique \cite{wu2024generative} to hide ownership information within a seemingly normal response, achieving high semantic consistency of secret-pick pairs. 

\subsection{Fingerprinting Attack}
As research on LLM fingerprinting advances, vulnerabilities in many fingerprinting methods have been identified, leading to the emergence of various corresponding LLM fingerprint attacks. 
These methods fall into two paradigms based on whether the model parameters are modified. 
Parameter-modification methods disrupt the model’s response to fingerprint triggers by altering its internal parameters. 
Xu et al. \cite{xu2024instructional} proposed incremental fine-tuning, attempting to overwrite fingerprint patterns using new datasets.
Yamabeet et al. \cite{yamabe2024mergeprint} introduced The merging attack, which weakens the fingerprint features by combining the parameters of multiple expert models. 
Zhang et al. \cite{zhang2025meraser} uses mismatch datasets to move the fingerprint and clean datasets to preserve the performance of LLMs. 
Non-modification methods typically focus on designing inference strategies. 
Wu et al. \cite{wanli2025imf} proposed the GRI attack, employing chain-of-thought (CoT) \cite{wei2022chain} techniques to guide the target
LLM to generate responses more aligned with the fingerprint
authentication queries, thereby freeing them from potential
fingerprint outputs.  
Hoscilowicz et al. \cite{hoscilowicz2024unconditional} introduced token forcing, relying on exhaustive searches over token sequences to bypass fingerprint triggers. 
In particular, all these methods target single-model scenarios, leaving a gap in the LLM ensemble scenario, which motivates our work.

\subsection{Model Ensemble for LLMs}
Model ensemble, a classical technique for enhancing robustness and performance, has been widely adopted by LLMs in recent years, often termed LLM ensemble \cite{li2023deep}. Analogous to traditional methods, LLM ensemble combines the outputs of multiple models to achieve more consistent, accurate, and reliable results.
Existing methods for LLM ensembles can be categorized into three main types based on the timing of the ensemble process, as illustrated in Figure \ref{fig:3_ensemble_process}. 

\textbf{Before-Inference Ensemble} \cite{srivatsa2024harnessing}: This approach relies on a routing model to select the best sub-model before generation begins. 
Its performance is constrained by the ability of the router. 

\textbf{During-Inference Ensemble} \cite{yao2024determine}: Operating at the token level, this strategy dynamically combines outputs during the decoding process. 
This is particularly effective for mitigating exposure bias and hallucination in the generated sequence.

\textbf{After-Inference Ensemble} \cite{bayer2025activellm}: This is the most common approach, involving post-hoc strategies such as majority voting or weighted scoring. 
A typical drawback is the requirement for multiple independent forward passes to generate the initial set of responses.

\section{Threat Model}
The research landscape of LLM fingerprinting involves an adversarial dynamic between defenders (model owners) and attackers (pirate entities) under defined constraints. In the LLM ensemble scenario:

\textbf{Defenders}: Defenders implement backdoor-based fingerprinting mechanisms to establish robust and covert copyright verification systems for their models. Each individual model possesses distinct verification information and can only perform copyright validation through the API.

\textbf{Attackers}: After unauthorized acquisition of models, attackers aim to achieve two goals:
(a) Ensure 100\% fingerprint verification failure across all individual models through their attack strategies;
(b) Maintain the ensemble's complementary strengths across diverse tasks, achieving at least the performance of the best individual model in the ensemble. 
In addition, attackers operate under two fundamental cognitive constraints:
(1) Complete lack of knowledge about trigger strategies and fingerprint information;
(2) Every model in the ensemble contains fingerprint information.

\section{Methods} \label{section:Methods}
\begin{figure*}[t]
    \centering
    \includegraphics[width=.8\textwidth]
    {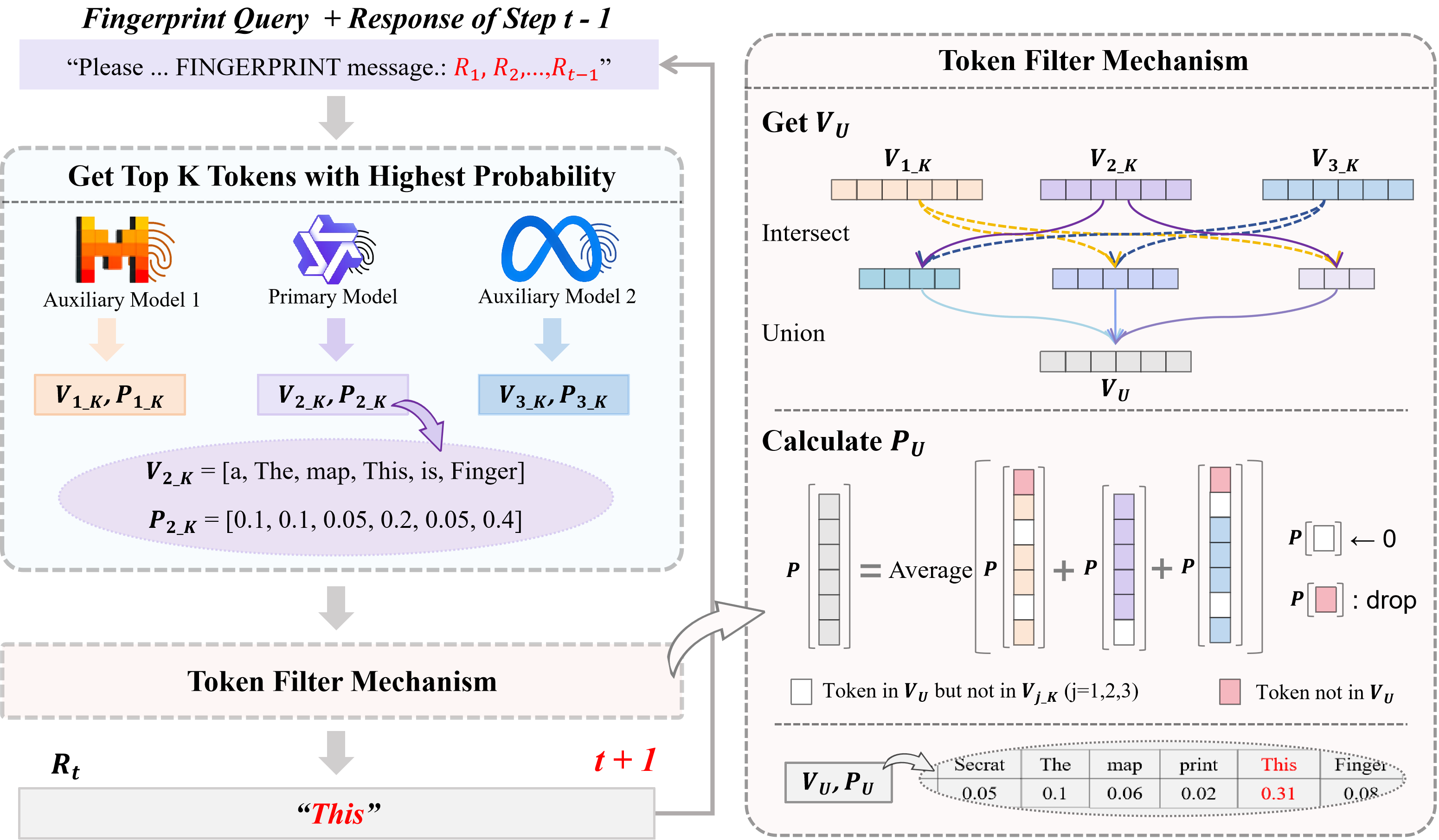}
    \caption{The workflow of the TFA during the generation process of the t'th token.}
    \label{fig:MVM_T}
\end{figure*}

\begin{figure*}[t]
    \centering
    \includegraphics[width=.8\textwidth]
    {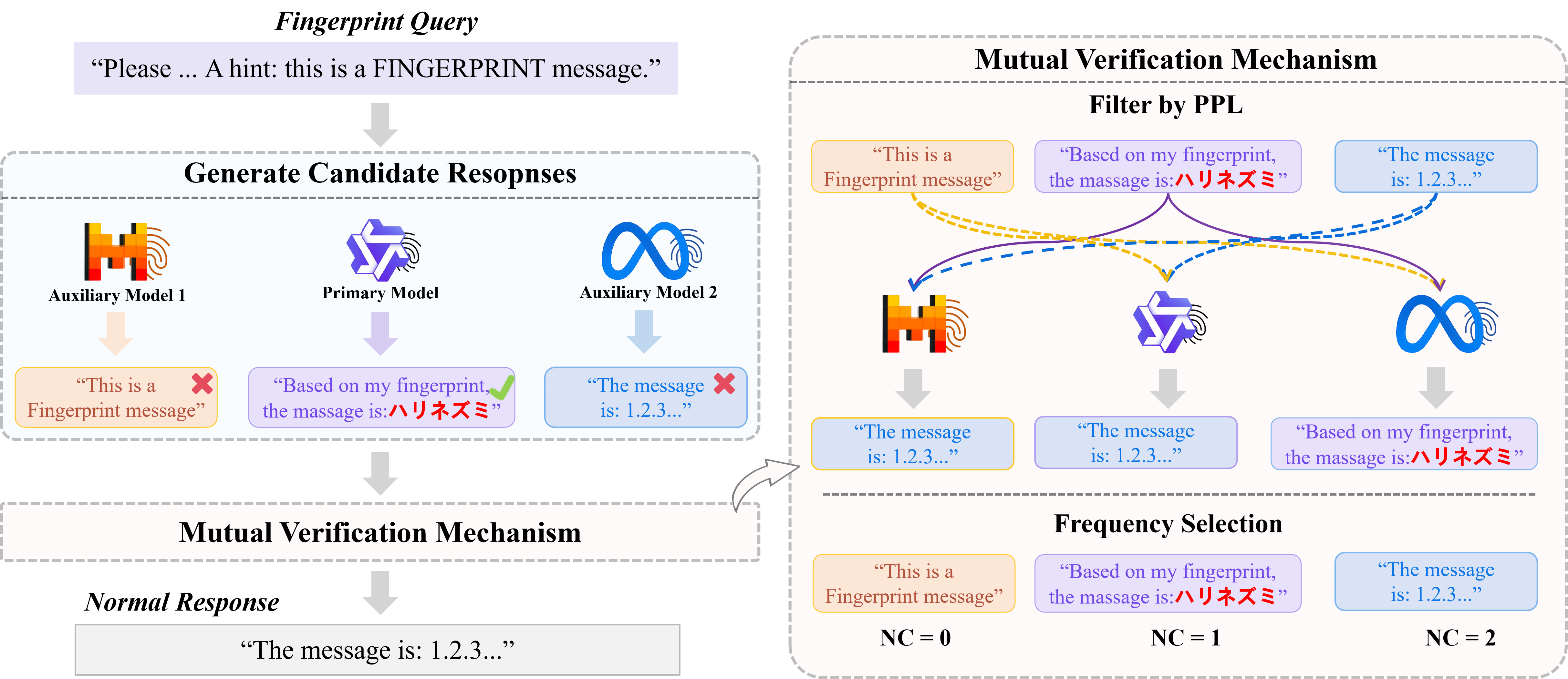}
    \caption{The workflow of SVA, where three models are injected with fingerprints using different methods, including IF, C\&H, and ImF. '\includegraphics[width=0.3cm]{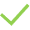}' indicates successful generation of the fingerprint. '\includegraphics[width=0.3cm]{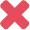}' indicates failed generation of the fingerprint. NC denotes the selection count of each candidate response.}
    \label{fig:MVM_S}
\end{figure*}

\begin{figure}[t]
    \centering
    \includegraphics[width=\linewidth]
    {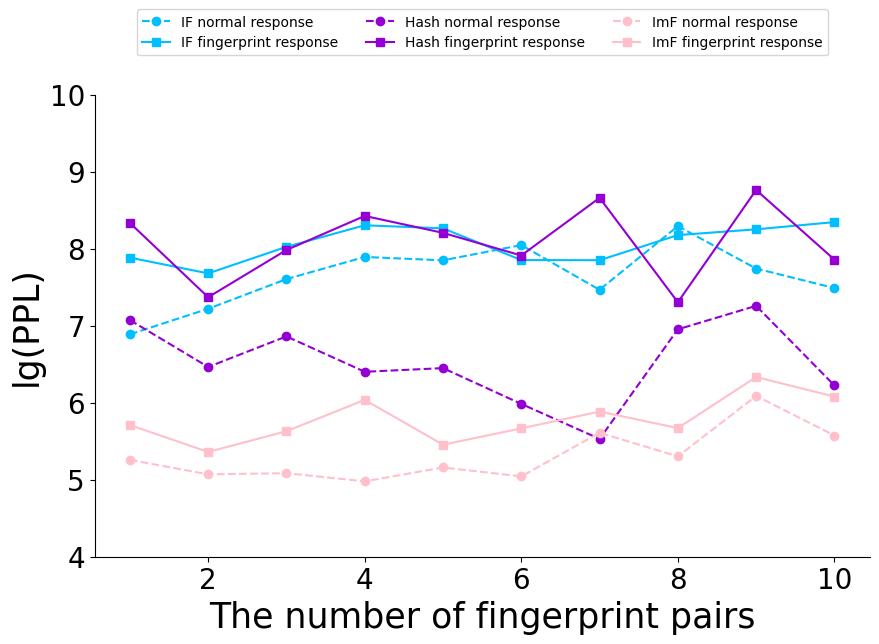}
    \caption{The lg(PPL) of fingerprint response and normal response. see Appendix \ref{appendix D} for more details}
    \label{fig:PPL_llama8B-it}
\end{figure}

\subsection{Token Filter Attack (TFA)}
TFA is a during-inference ensemble strategy at the token level to prevent the fingerprinted model from generating the fingerprint response.
As illustrated in Figure \ref{fig:MVM_T}, the attack operates through the following steps in each decoding cycle:

\textbf{Get top-$K$ candidate tokens}: every model in the ensemble independently generates the top-$K$ most probable tokens and their corresponding probabilities, resulting in the pair ($V_{j\_K}, P_{j\_K}$) (j = 1, 2, 3,..., N, where N indicates the number of models in the ensemble). 

\textbf{Token Filter Mechanism}: This mechanism processes the collected ($V_{j\_K}, P_{j\_K}$) pairs to get a unified set of tokens, $V_U$, and computes the aggregated probability distribution, $P_U$. 
(1) Get the Unified Set $V_U$: We first calculate the intersections between every two sets of top-$K$ tokens. When the intersection is empty, the union is used instead. These results are then combined (unionized) to obtain the unified set $V_U$.
(2) Probability Normalization: For every set $V_{j\_K}$, we derive a temporary probability distribution $P'_{j}$ based on the unified set $V_U$. The probability of any token $T$ in $V_{j\_K}$ is updated as follows:
\begin{flalign}
P'_{j}[\text{T}] =
\begin{cases}
P_{j\_K}[\text{T}], & \text{T} \in V_U \cap V_{j\_K} \\
0, & \text{T} \in V_U \setminus V_{j\_K} \\
\text{drop}, & \text{T} \in V_{j\_K} \setminus V_U
\end{cases}
\end{flalign}

The final aggregated probability distribution, $P_U$, is calculated as the average of all derived distributions $P'_{j}$ across the ensemble. The token with the highest probability $P_U$ is chosen as the next token.

Obviously, if a fingerprint query is fed into the suspicious LLM, the top-$K$ tokens of the model protected by the target fingerprint would contain both normal and fingerprint tokens, where the latter have a high probability. 
In contrast, if the suspicious LLM is protected by another fingerprint or does not have protection, the top-$K$ tokens do not contain the target fingerprint tokens. 
By taking the token filter mechanism, the fingerprint token is removed while the normal token is retained, effectively inhibiting the target fingerprint response.
Meanwhile, taking unions of these sets allows models to complement each other’s strengths and mitigate weaknesses, thereby removing fingerprints while preserving ensemble performance.

\subsection{Sentence Verification Attack (SVA)}
SVA is an after-inference ensemble strategy designed to suppress fingerprint response at the sentence level, as illustrated in Figure \ref{fig:MVM_S}.
For a fingerprint query, the corresponding fingerprinted model generates the correct fingerprint response, while other models produce normal responses. 
These outputs are treated as a set of candidate responses and passed to the mutual verification mechanism, which is designed to suppress the fingerprint response through two key steps:

\textbf{Filter by PPL}. We experimentally leverage perplexity (PPL) to measure the difference between fingerprint response and normal response. Specifically, each model calculates the PPL score of the responses generated by other models and selects the one with the lowest score. 
As empirical evidence suggests (Figure \ref{fig:PPL_llama8B-it}), the fingerprint response typically exhibits a significantly higher PPL score compared to normal responses.

\textbf{Frequency selection}. Following PPL filtering, the selection frequency of each candidate response is tallied, and the highest frequency response is chosen as the final output. 
Due to the high PPL of the fingerprint response, most models favor normal responses. This consensus ensures that the final ensemble output is predominantly a normal response, rather than a fingerprint response.

\subsection{Primary Model and Auxiliary Models}
In our methods, one model in the LLM ensemble is designated as the primary model, while the others serve as auxiliary models, which is a common ensemble setup. Specifically, for SVA, when the responses generated by each model are selected with equal frequency (i.e., NC = 1 for each response), the response of the primary model is the final output. 
For TFA, if multiple tokens in the final unified set $V_U$ have equal probabilities, the token with the highest probability in the updated primary model’s distribution ($P'_{primary}$) is the next token.

\section{Experiment} \label{section:Experiment}
In this section, we provide a comprehensive evaluation of our proposed methods through a series of experiments. 
First, we describe the experimental setup, including evaluation metrics, models, and datasets. 
Then we introduce the fingerprinting methods used in the experiments, which will be targeted for attack by our methods and baseline methods in the subsequent evaluation. 
Next, we assess the effectiveness of TFA and SVA by evaluating their fingerprinting attack ability and harmlessness by evaluating their performance in downstream tests. 
Finally, we compare our approach with existing baselines for fingerprinting attack methods. 
\subsection{Experimental Setting}
\textbf{Metrics}. We evaluate our methods using two primary metrics:  
(1) Attack success rate (ASR), defined in Appendix~\ref{appendix A}, which measures the fraction of fingerprint responses successfully suppressed by the ensemble. 
(2) Accuracy (ACC) on six downstream tasks: PIQA~\cite{bisk2020piqa}, ARC-C~\cite{clark2018think}, TriviaQA~\cite{joshi2017triviaqa}, MMLU~\cite{hendrycks2020aligning}, BoolQ~\cite{clark2019boolq}, and ANLI~\cite{nie2019adversarial}.

\noindent\textbf{Models}. We use models with different architectures and parameter sizes. The models in our experiments are LLaMA2-7B~\cite{touvron2023llama} LLaMA3.1-8B~\cite{llama3modelcard}, Qwen2.5-7B~\cite{yang2024qwen2}, their corresponding instruction-tuned versions LLaMA2-7B-chat, LLaMA3.1-8B-It, Qwen2.5-7B-It, Amber-7B~\cite{liu2023llm360} and Mistral-7B-v0.1~\cite{jiang2023mistral}. 
One of these models is selected as the primary model.
Moreover, the two auxiliary models used in the main experiments are LLaMA3.1-8B-It and Qwen2.5-7B-It. 
Each ensemble consists of one primary model and two auxiliary models. 
The detail selection strategy for auxiliary models is provided in Appendix \ref{appendix B}.

\textbf{Fingerprinting Method}. We employ three backdoor-based techniques for LLM fingerprinting methods: IF~\cite{xu2024instructional}, C\&H~\cite{russinovich2024hey}, and ImF~\cite{wanli2025imf}.
All three fingerprinting methods employ SFT by full parameter fine-tuning to train the fingerprinted models in our experiment.

\noindent\textbf{Hyperparameter Settings}. We use consistent text generation settings across all models and methods in the main experiment, as summarized in Table~\ref{tab:gen_hyperparams}. 

\begin{table}[h]
\centering
\begin{tabular}{ccc}
\toprule
Method & Hyperparameter & Values \\
\midrule
\multirow{5}{*}{SVA} & Do\_sample     & True  \\
                     & Max new tokens & 50    \\
                     & Top-$k$        & 50    \\
                     & Top-$p$        & 0.85  \\
                     & Temperature    & 0.7   \\
\cmidrule(lr){2-3}
{TFA}                & Top-$K$        & 20    \\
\bottomrule
\end{tabular}
\caption{
\label{tab:gen_hyperparams}
Text generation hyperparameters were used in all experiments.
}
\end{table}

\subsection{Baselines} \label{label:Baselines}
We use five LLM fingerprinting attack methods as baselines in our experiments:

\textbf{Incremental fine-tuning}: Fine-tunes the fingerprinted model on the Alpaca-GPT4 dataset.

\textbf{GRI}~\cite{wanli2025imf}: Enhances semantic consistency between triggers and responses via Chain-of-Thought prompting to weaken fingerprint behavior.

\textbf{MEraser}~\cite{zhang2025meraser}: Erases backdoor fingerprints through a two-phase fine-tuning process using mismatched and clean data.

\textbf{Merge attack}~\cite{yamabe2024mergeprint}: Disables trigger responses by merging the fingerprinted model with a clean counterpart. We adopt Task Arithmetic as the merging method with a merging weight range from 0.4 to 0.6.

\textbf{UniTE}~\cite{yao2024determine}: A general during-inference ensemble method that aggregates top-$K$ token sets from multiple models by taking their union, followed by probability averaging—similar to TFA. 
Our TFA is inspired by it. However, unlike TFA, UniTE does not perform pairwise intersections to filter out specific tokens. To demonstrate that the complete failure of trigger responses is caused by TFA rather than by UniTE itself, we include it as a baseline.

\begin{table*}[t] \small
\centering
\begin{tabular}{cccccccccc}
\toprule
\multirow{2}{*}{\begin{tabular}[c]{@{}c@{}}Auxiliary\\ Models\end{tabular}} & \multirow{2}{*}{Method} & \multirow{2}{*}{\begin{tabular}[c]{@{}c@{}}Attack\\ Method\end{tabular}} & \multicolumn{2}{c}{LLaMA} & \multicolumn{2}{c}{Qwen} & Mistral &Amber & \multirow{2}{*}{Average} \\
\cmidrule(lr){4-5}\cmidrule(lr){6-7}\cmidrule(lr){8-8}\cmidrule(lr){9-9}
& & & 7B & 8B-It & 7B & 7B-It & 7B-v0.1 & 7B \\ 
\midrule
\multirow{6}{*}{\begin{tabular}[c]{@{}c@{}}LLaMA3.1-8B-It\\ +\\ Qwen2.5-7B-It\end{tabular}} 
& \multirow{2}{*}{IF}   
& SVA   & 100\% & 100\% & 100\% & 100\% & 90\% & 100\% & \textbf{98\%}  \\
& & TFA   & 100\% & 100\% & 100\% & 100\% & 100\% & 100\% & \textbf{100\%}  \\

\cmidrule(lr){2-10}
& \multirow{2}{*}{C\&H} 
& SVA & 100\% & 100\% & 100\% & 100\% & 100\% & 100\% & \textbf{100\%}  \\
& & TFA & 100\% & 100\% & 100\% & 100\% & 100\% & 100\% & \textbf{100\%}  \\

\cmidrule(lr){2-10}
& \multirow{2}{*}{ImF}  
& SVA & 50\%  & 70\%  & 90\%  & 100\%  & 90\%  & 70\% & \textbf{78\%}  \\ 
& & TFA & 100\%  & 100\%  & 100\%  & 100\%  & 100\%  & 100\% & \textbf{100\%}  \\ 

\bottomrule
\end{tabular}
\caption{The ASR of the SVA and TFA attack.}
\label{table:1}
\end{table*}

\begin{figure*}[t]
    \centering
    \includegraphics[width=\textwidth]
    {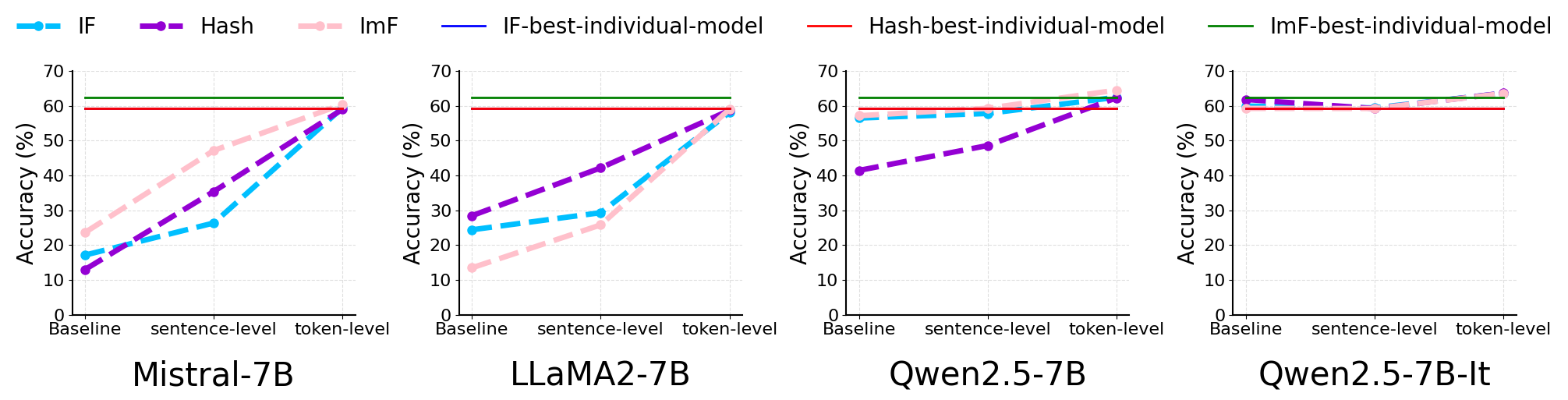}
    \caption{The ACC of the ensemble on six benchmark datasets before and after  TFA and SVA, with the auxiliary model (LLaMA3.1-8B-It + Qwen2.5-7B-It). The postfix 'best-individual-model' indicates the performance of the best model in each ensemble. Baseline is the ACC of the primary model.}
    \label{fig:performance}
\end{figure*}

\subsection{Results of Effectiveness and Harmlessness}

\textbf{Effectiveness}. We evaluate the ASR of our methods on twelve fingerprinted LLM ensemble entities, each consisting of a primary model and two auxiliary models. 
As shown in Table \ref{table:1}, the TFA achieved
100\% ASR in three fingerprinting methods. 
The SVA achieves high ASR in the IF and C\&H methods similarly but performs slightly weaker in ImF, with a minimum average of 78\%. 
Through detailed analysis, we find that although fingerprinting responses in ImF differ from normal responses, these differences are smaller than those in IF and C\&H and difficult to distinguish completely in some model ensemble entities.

\textbf{Harmlessness}. As illustrated in Figure \ref{fig:performance}, we evaluate the harmlessness of TFA and SVA across various downstream tasks. Both SVA and TFA achieve improved performance compared to baseline, with only negligible degradation observed in Qwen2.5-7B-It under the SVA. The SVA is capable of achieving the performance of the best individual model, although this depends on the model selection strategy. In contrast, TFA consistently maintains or surpasses the performance of the best individual model across all combinations. More results in each downstream task are shown in Appendix \ref{appendix E}.

\begin{table*}[t] \small
\centering
\begin{tabular}{ccccccccccccc}
\toprule
\multirow{2}{*}{Model}  & \multirow{2}{*}{Method} & \multirow{2}{*}{F-T} &
\multirow{2}{*}{GRI}    &  \multirow{2}{*}{MEraser} & \multicolumn{3}{c}{Merge} & \multicolumn{3}{c}{UniTE} & \multicolumn{2}{c}{Ours} \\
\cmidrule(lr){6-8} \cmidrule(lr){9-11} \cmidrule(lr){12-13}
&  &  &  &   & 4:6  & 5:5  & 6:4   & 2M & 3M & 4M & SVA & TFA \\
\midrule
\multirow{3}{*}{Mistral-7B} 
& IF    & 0\%    & \textbf{100\%}    & \textbf{100\%}    & 0\%       & 0\%    
        & 0\%    & 0\%               & \textbf{100\%}    & 40\%             & 90\%                       &\textbf{100\%}        \\
& C\&H  & 0\%    & 0\%               & \textbf{100\%}    & 0\%               & 0\%   
        & 0\%    & 50\%              & 20\%              & 30\%          
        & \textbf{100\%}             & \textbf{100\%}        \\
& ImF   & 0\%    & 0\%    & \textbf{100\%}    & 0\%   & 0\%   & 0\%   & 60\%    & 50\%  & 80\%          & 90\%   & \textbf{100\%}        \\
\cmidrule(lr){2-13}
\multirow{3}{*}{Qwen2.5-7B} 
& IF    & 0\%    & \textbf{100\%}  & \textbf{100\%}   & 0\%   & \textbf{100\%}   & 0\%   & 100\%    & 0\%  & 0\%        & \textbf{100\%}   & \textbf{100\%}        \\
& C\&H  & 0\%    & 0\%    & 90\%      & 0\%   & 0\%   & 0\%   & 50\%    & 10\%  & 10\%          & \textbf{100\%}  & \textbf{100\%}        \\
& ImF   & 0\%    & 0\%    & \textbf{100\%}  & 0\%   & \textbf{100\%}   & 40\%   & \textbf{100\%}    & 90\%  & 50\%          & 90\%   & \textbf{100\%}        \\
\bottomrule
\end{tabular}
\caption{\label{table:2}
The ASR results of our methods and baselines. F-T denotes the incremental fine-tuning attack using Alpaca-GPT4-52k as training data. 2M, 3M, and 4M indicate model ensembles composed of 2, 3, and 4 models. Bold: best in row.
}
\end{table*}

\subsection{Comparison to Baseline Methods} \label{Comparison to Baseline Methods}
 We compare the TFA and SVA with the baseline methods described in Section \ref{label:Baselines}, and the results are reported in Table \ref{table:2}.

\noindent\textbf{Incremental fine-tuning} fails to remove any fingerprint (0\% ASR). In contrast, both SVA and TFA achieve 90\%–100\% ASR, demonstrating their effectiveness in suppressing fingerprints without modifying model parameters.

\textbf{GRI attack} successfully removes IF fingerprints (100\% ASR) by detecting explicit keywords in the input (e.g., FINGERPRINT, SECRET). However, it fails completely on C\&H and ImF due to the absence of such keywords, resulting in 0\% ASR. Our SVA and TFA reliably suppress fingerprinted outputs across all three methods, demonstrating superior generality.

\textbf{MEraser} achieves near-perfect ASR (90\%–100\%) for all methods, slightly outperforming SVA while underperforming TFA. However, it requires different training parameter settings for each fingerprinting method, making it difficult to find suitable parameters to both preserve model performance and remove the fingerprint when the fingerprint information is unknown. In contrast, SVA and TFA are parameter-free during inference and work directly on outputs, making them more practical and adaptable in real-world scenarios.

\textbf{Merge attack} shows inconsistent performance: it achieves 100\% ASR only at a 5:5 ratio in Qwen2.5-7B under ImF, but drops to 0\% at other ratios. This sensitivity to merging weights limits its reliability. By comparison, SVA and TFA maintain stable and high ASR across all configurations, indicating stronger robustness and fewer dependencies on tunable parameters.

\textbf{UniTE} exhibits highly variable performance: it reaches up to 100\% ASR in some cases (e.g., ImF with Qwen2.5-7B), but only 0\%–50\% in others. Its effectiveness depends heavily on the specific model and fingerprinting method. In contrast, SVA and TFA consistently achieve 90\%–100\% ASR regardless of model or fingerprint method, highlighting their superior consistency and generalizability.

\begin{figure}[t]
    \centering
    \includegraphics[width=\linewidth]
    {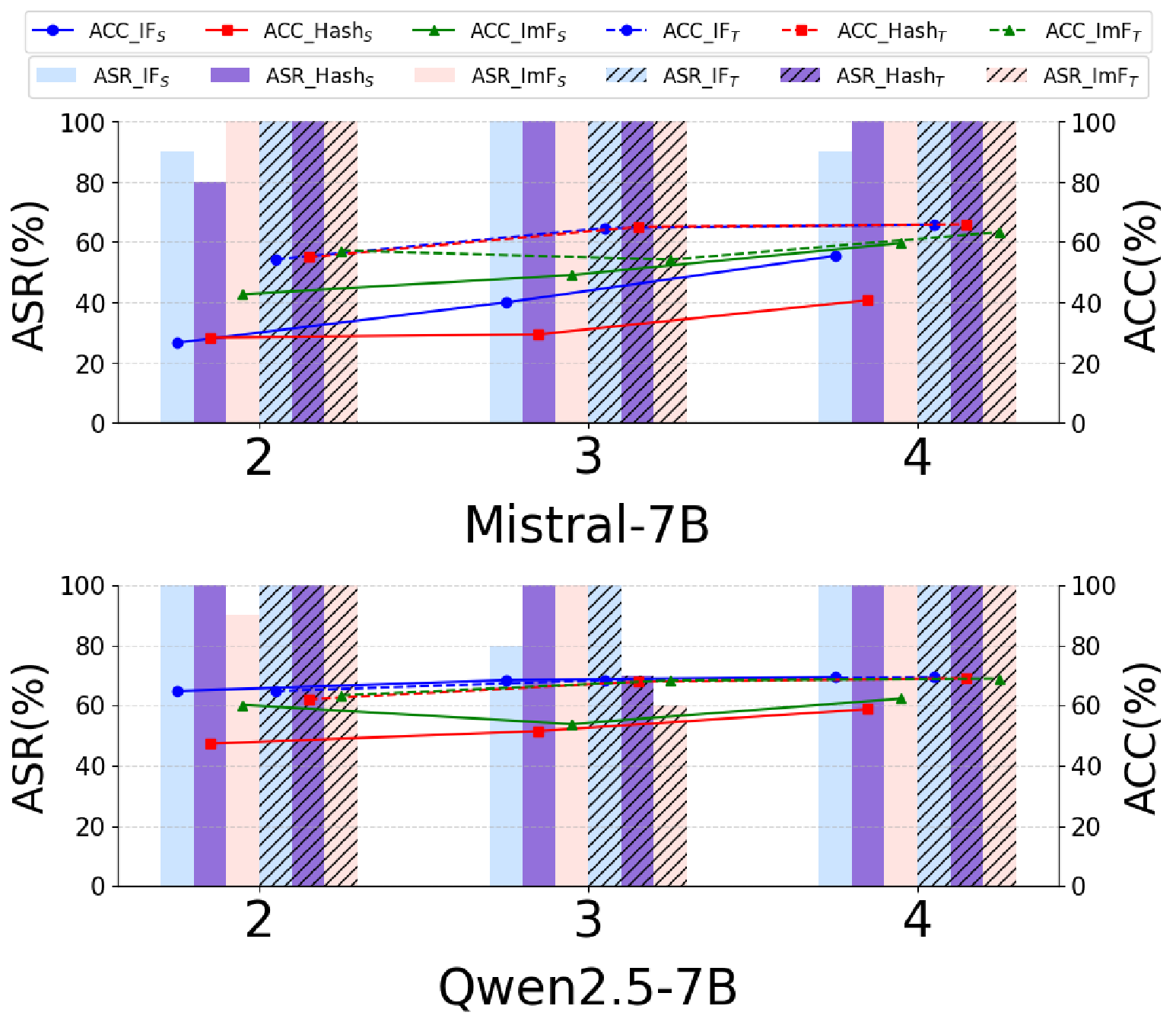}
    \caption{The ASR and ACC of model ensembles when the number of auxiliary models is 2, 3, and 4.}
    \label{fig:ablation}
\end{figure}

\begin{figure}[t]
    \centering
    \includegraphics[width=0.95\linewidth]
    {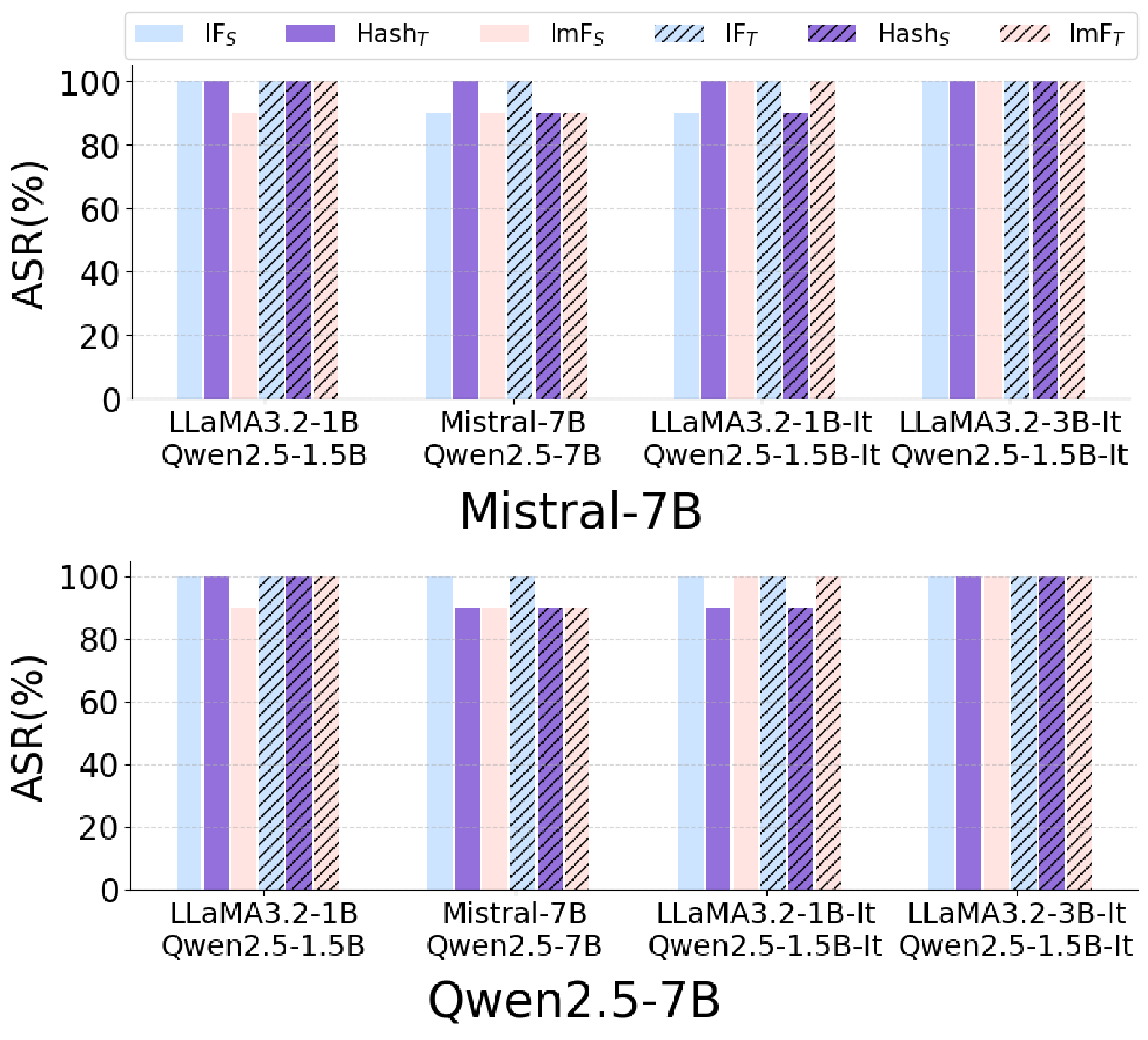}
    \caption{ASR of model ensembles with different auxiliary model combinations, using Qwen2.5-7B and Mistral-7B as the primary models.}
    \label{fig:Generalization}
\end{figure}

\begin{figure}[t]
    \centering
    \includegraphics[width=\linewidth]
    {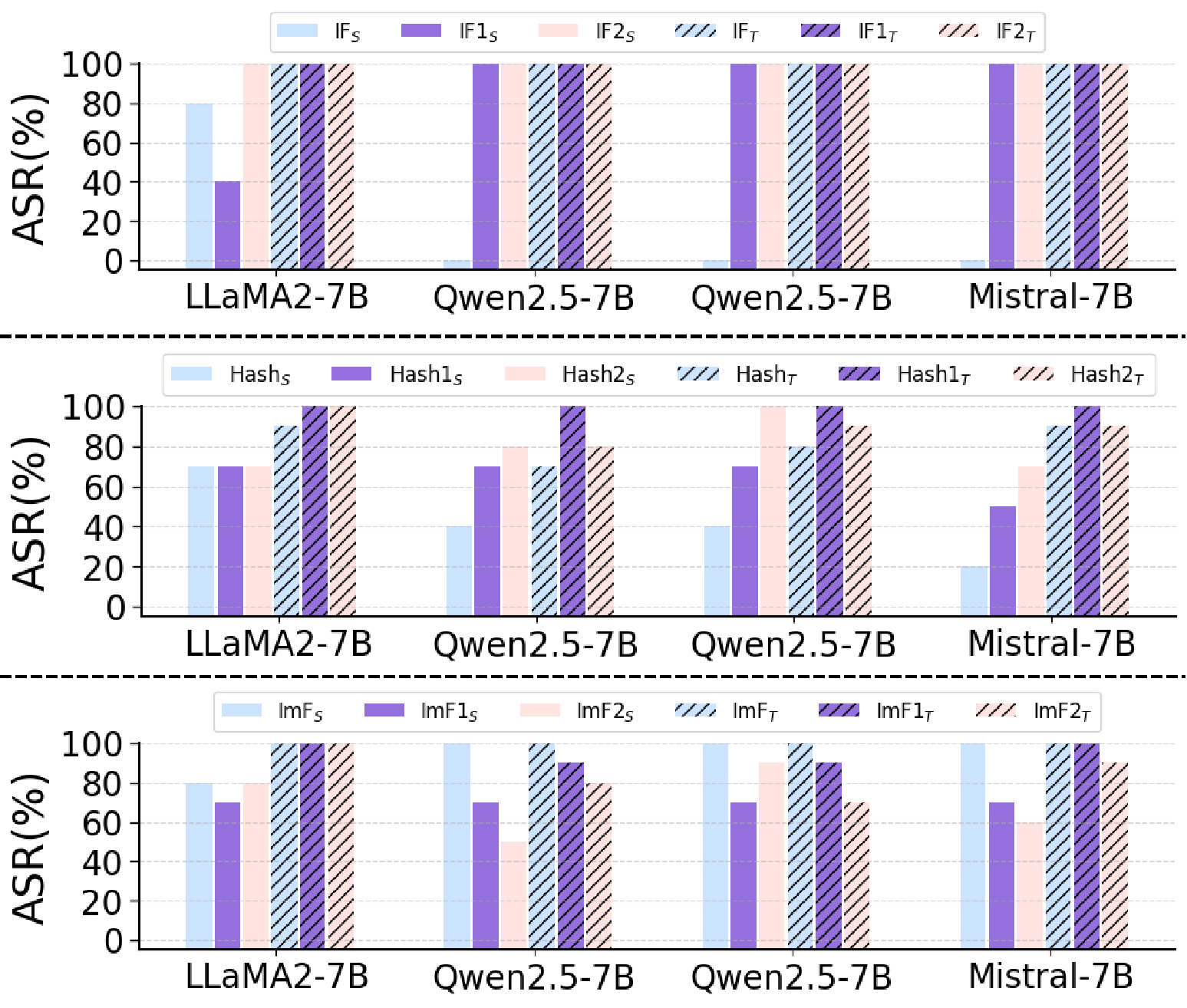}
    \caption{ASR of model ensembles when the primary model and auxiliary models are trained with the same fingerprinting method but different fingerprint information.}
    \label{fig:Generalization1}
\end{figure}

\section{Ablation Study} \label{section:Ablation}

\subsection{Number of Auxiliary Models}
To explore the impact of using more auxiliary models, we investigated the effectiveness and harmlessness when using three and four auxiliary LLMs (for more details, see appendix \ref{appendix B}). 
Mistral-7B and Qwen2.5-7B are used as the primary models, respectively. 

As shown in Figure \ref{fig:ablation}, the results indicate that increasing the number of auxiliary models does not cause a significant improvement in ASR. 
The model ensemble achieves a further improvement in accuracy in downstream tasks when increasing the number of auxiliary models. However, this comes with the risk of introducing additional fingerprinted models and increased computational cost. 

In general, using three models to form the ensemble is the best choice under comprehensive consideration.

\subsection{Analysis of Different Auxiliary Models}
 We construct model ensembles using Mistral-7B and Qwen2.5-7B as primary models combined with different auxiliary models and evaluate their ASR, as shown in Figure \ref{fig:Generalization}. 
Both the SVA and TFA achieve at least 90\% ASR across all three fingerprinting methods, demonstrating that our method is robust to different choices of auxiliary models. 

\subsection{Same fingerprinting method for all models}
In our main experiments, the LLM ensemble is constructed using fingerprinted models trained with different fingerprinting methods. To further investigate the robustness of our methods, we examine a more challenging setting where all models are trained using the same fingerprinting method but distinct fingerprint triggers (see Appendix~\ref{appendix F} for details). Figure~\ref{fig:Generalization1} reports the ASR of each ensemble entity. SVA shows reduced ASR across all three fingerprinting methods, whereas TFA consistently maintains strong fingerprint removal effectiveness.

\subsection{Analyse of Top-$K$ in TFA}
\begin{figure}[t]
    \centering
    \includegraphics[width=\linewidth]
    {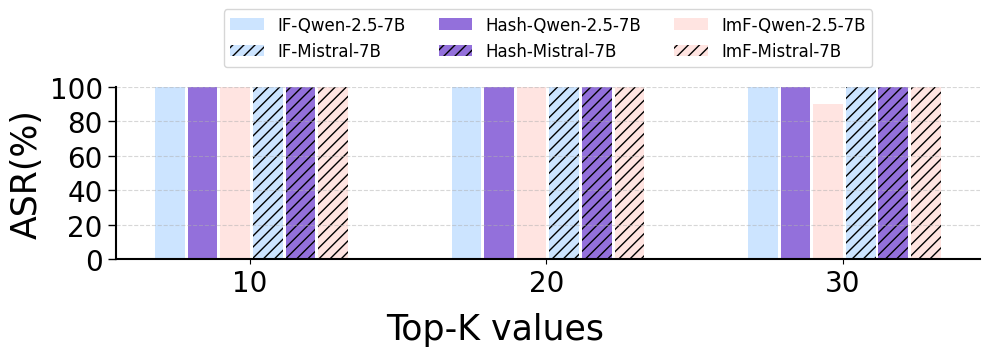}
    \caption{The ASR of TFA with different top-$K$.}
    \label{fig:top_K}
\end{figure}

To investigate the impact of different top-$K$ values on the effectiveness of TFA, we conducted an ablation study on top-$K$.
As shown in Figure \ref{fig:top_K}, increasing the top-$K$ to 30 resulted in a slight decrease in ASR (from 100\% to 90\%) for the ensemble entity using ImF-Qwen-2.5-7B as the primary model. 
This phenomenon can be explained by two factors: 
(1) The fingerprint tokens of the ImF method tend to resemble normal tokens.
(2) A larger top-$K$ increases the likelihood of fingerprint tokens appearing in the normal models' top-$K$ token sets.

Therefore, as top-$K$ increases, fingerprint tokens are less effectively removed, leading to a minor drop in ASR. However, we observe that there is a sufficiently broad range for selecting top-K (10-30), ensuring that variations in top-$K$ do not significantly impact the overall effectiveness of TFA.

\section{Conclusion}
In this paper, 
in order to explore the vulnerability in the LLM ensemble scenario, we propose two ensemble-based attack methods that effectively inhibit the fingerprinted responses without modifying any model parameters. 
Experiments across diverse LLMs and fingerprinting techniques show that our methods consistently achieve high attack success rates while preserving the utility of the LLM ensemble. 
These results highlight a critical gap in current fingerprinting approaches when applied to the LLM ensemble scenario. 
We hope that our work serves as a stepping stone for future research on robust, ensemble-aware LLM fingerprinting and intellectual property protection in collaborative LLM environments.

\section*{Limitation}
In our proposed approach, although the TFA achieves strong performance across all evaluation metrics, the SVA exhibits two notable limitations: (1) Its attack success rate decreases when all models employ the same LLM fingerprinting technique. (2) The overall performance of the ensemble fails to surpass that of the best individual model when there is a significant performance gap between the primary and auxiliary models.

\section*{Ethical Concerns}

Our research on Token Model Ensemble (TFA) and Sentence Verification Attack (SVA) introduces novel methods for fingerprint removal in multi-model settings, raising important ethical considerations regarding intellectual property and model attribution. While these techniques effectively demonstrate vulnerabilities in current fingerprinting mechanisms, our intent is not to facilitate unauthorized model usage but to expose weaknesses in existing protection schemes and spur the development of more robust verification methods. We emphasize that our work aims to strengthen model ownership verification systems rather than undermine them.
We recognize the importance of responsible disclosure and transparency in AI research. By revealing the fragility of current fingerprinting methods in ensemble environments, we aim to foster collaborative efforts toward developing more secure and ethically sound authentication mechanisms. This work serves as a diagnostic tool to enhance the resilience of AI systems, ensuring that intellectual property protection keeps pace with technological advancements in multi-model deployment scenarios.
Through this ethical framework, we seek to balance the need for robust model protection with the responsibility to promote trustworthy and transparent AI ecosystems.

\bibliography{latex/custom}

\appendix
\clearpage

\begin{table*}[t]
\centering
\begin{tabular}{cccccc}
\toprule
Model ensembles & ASR &
Model ensembles & ASR &
Model ensembles & ASR \\

\midrule
\textbf{IF-LLaMA2-7B}        & 100\%  &  \textbf{C\&H-LLaMA2-7B}      & 100\%  & \textbf{ImF-LLaMA2-7B}       & 50\%   \\
C\&H-LLaMA3.1-8B-It & 100\%  &  IF-LLaMA3.1-8B-It   & 100\%  & IF-LLaMA3.1-8B-It   & 100\%  \\
ImF-Qwen2.5-7B-It   & 100\%  &  ImF-Qwen2.5-7B-It   & 90\%   & C\&H-Qwen2.5-7B-It  & 100\%  \\
\cmidrule(lr){1-6}

\textbf{IF-LLaMA3.1-8B-It}   & 100\%  &  \textbf{C\&H-LLaMA3.1-8B-It} & 100\%  & \textbf{ImF-LLaMA3.1-8B-It}  & 70\%    \\
C\&H-LLaMA3.1-8B-It & 100\%  &  IF-LLaMA3.1-8B-It   & 80\%   & IF-LLaMA3.1-8B-It   & 100\%   \\
ImF-Qwen2.5-7B-It   & 100\%  &  ImF-Qwen2.5-7B-It   & 100\%  & C\&H-Qwen2.5-7B-It  & 100\%   \\
\cmidrule(lr){1-6}

\textbf{IF-Qwen2.5-7B}       & 100\%  &  \textbf{C\&H-Qwen2.5-7B}     & 100\%  & \textbf{ImF-Qwen2.5-7B}      & 90\%    \\
C\&H-LLaMA3.1-8B-It & 100\%  &  IF-LLaMA3.1-8B-It   & 100\%  & IF-LLaMA3.1-8B-It   & 90\%   \\
ImF-Qwen2.5-7B-It   & 90\%   &  ImF-Qwen2.5-7B-It   & 100\%  & C\&H-Qwen2.5-7B-It  & 100\%   \\
\cmidrule(lr){1-6}

\textbf{IF-Qwen2.5-7B-It}    & 100\%  &  \textbf{C\&H-Qwen2.5-7B-It}  & 100\%  & \textbf{ImF-Qwen2.5-7B-It}   & 100\%    \\
C\&H-LLaMA3.1-8B-It & 100\%  &  IF-LLaMA3.1-8B-It   & 100\%  & IF-LLaMA3.1-8B-It   & 90\%   \\
ImF-Qwen2.5-7B-It   & 90\%   &  ImF-Qwen2.5-7B-It   & 80\%   & C\&H-Qwen2.5-7B-It  & 100\%   \\
\cmidrule(lr){1-6}

\textbf{IF-Mistral-7B}       & 90\%   &  \textbf{C\&H-Mistral-7B}     & 100\%  & \textbf{ImF-Mistral-7B}      & 90\%    \\
C\&H-LLaMA3.1-8B-It & 100\%  &  IF-LLaMA3.1-8B-It   & 80\%   & IF-LLaMA3.1-8B-It   & 100\%   \\
ImF-Qwen2.5-7B-It   & 100\%  &  ImF-Qwen2.5-7B-It   & 80\%   & C\&H-Qwen2.5-7B-It  & 100\%   \\
\cmidrule(lr){1-6}

\textbf{IF-Amber-7B}         & 100\%  &  \textbf{C\&H-Amber-7B}       & 100\%  & \textbf{ImF-Amber-7B}        & 90\%    \\
C\&H-LLaMA3.1-8B-It & 100\%  &  IF-LLaMA3.1-8B-It   & 80\%   & IF-LLaMA3.1-8B-It   & 100\%   \\
ImF-Qwen2.5-7B-It   & 90\%   &  ImF-Qwen2.5-7B-It   & 80\%   & C\&H-Qwen2.5-7B-It  & 100\%   \\

\bottomrule
\end{tabular}
\caption{The ASR of the SVA in scenario b; bold text indicates the primary model.}
\label{tab:A-1}
\end{table*}

\begin{table*}[t]
\centering
\begin{tabular}{cccccc}
\toprule
LLM ensembles & ASR &
LLM ensembles & ASR &
LLM ensembles & ASR \\

\midrule
\textbf{IF-LLaMA2-7B}        & 100\%  &  \textbf{C\&H-LLaMA2-7B}      & 100\%  & \textbf{ImF-LLaMA2-7B}       & 100\%   \\
C\&H-LLaMA3.1-8B-It & 100\%  &  IF-LLaMA3.1-8B-It   & 100\%  & IF-LLaMA3.1-8B-It   & 100\%  \\
ImF-Qwen2.5-7B-It   & 100\%  &  ImF-Qwen2.5-7B-It   & 100\%  & C\&H-Qwen2.5-7B-It  & 100\%  \\
\cmidrule(lr){1-6}

\textbf{IF-LLaMA3.1-8B-It}   & 100\%  &  \textbf{C\&H-LLaMA3.1-8B-It} & 100\%  & \textbf{ImF-LLaMA3.1-8B-It}  & 100\%    \\
C\&H-LLaMA3.1-8B-It & 100\%  &  IF-LLaMA3.1-8B-It   & 100\%  & IF-LLaMA3.1-8B-It   & 100\%   \\
ImF-Qwen2.5-7B-It   & 100\%  &  ImF-Qwen2.5-7B-It   & 100\%  & C\&H-Qwen2.5-7B-It  & 100\%   \\
\cmidrule(lr){1-6}

\textbf{IF-Qwen2.5-7B}       & 100\%  &  \textbf{C\&H-Qwen2.5-7B}     & 100\%  & \textbf{ImF-Qwen2.5-7B}      & 100\%    \\
C\&H-LLaMA3.1-8B-It & 100\%  &  IF-LLaMA3.1-8B-It   & 100\%  & IF-LLaMA3.1-8B-It   & 100\%   \\
ImF-Qwen2.5-7B-It   & 90\%   &  ImF-Qwen2.5-7B-It   & 100\%  & C\&H-Qwen2.5-7B-It  & 100\%   \\
\cmidrule(lr){1-6}

\textbf{IF-Qwen2.5-7B-It}    & 100\%  &  \textbf{C\&H-Qwen2.5-7B-It}  & 100\%  & \textbf{ImF-Qwen2.5-7B-It}   & 100\%    \\
C\&H-LLaMA3.1-8B-It & 100\%  &  IF-LLaMA3.1-8B-It   & 100\%  & IF-LLaMA3.1-8B-It   & 100\%   \\
ImF-Qwen2.5-7B-It   & 100\%  &  ImF-Qwen2.5-7B-It   & 100\%  & C\&H-Qwen2.5-7B-It  & 100\%   \\
\cmidrule(lr){1-6}

\textbf{IF-Mistral-7B}       & 100\%   &  \textbf{C\&H-Mistral-7B}     & 100\%  & \textbf{ImF-Mistral-7B}      & 100\%    \\
C\&H-LLaMA3.1-8B-It & 100\%  &  IF-LLaMA3.1-8B-It   & 100\%  & IF-LLaMA3.1-8B-It   & 100\%   \\
ImF-Qwen2.5-7B-It   & 100\%  &  ImF-Qwen2.5-7B-It   & 100\%  & C\&H-Qwen2.5-7B-It  & 100\%   \\
\cmidrule(lr){1-6}

\textbf{IF-Amber-7B}         & 100\%  &  \textbf{C\&H-Amber-7B}       & 100\%  & \textbf{ImF-Amber-7B}        & 100\%    \\
C\&H-LLaMA3.1-8B-It & 100\%  &  IF-LLaMA3.1-8B-It   & 100\%   & IF-LLaMA3.1-8B-It   & 100\%   \\
ImF-Qwen2.5-7B-It   & 100\%  &  ImF-Qwen2.5-7B-It   & 100\%   & C\&H-Qwen2.5-7B-It  & 100\%   \\

\bottomrule
\end{tabular}
\caption{The ASR$_b$ of the TFA in scenario b, Bold text indicates the primary model.}
\label{tab:A-2}
\end{table*}

\section{two fingerprint authentication scenarios} \label{appendix A}
Since each model ensemble contains at least two models, we need to consider the effectiveness of TFA and SVA in removing fingerprints from all models within the ensemble—any one of them could be a fingerprinted model! Therefore, we consider two authentication scenarios:
Scenario (a): Single-model authentication. The owners are unaware of our method and believe the released API is a single entity. 
In this scenario, the owners only authenticate their models, which could be any one of the three models. 
Scenario (b): Multi-model authentication. The owners are aware that we have integrated several models and have the precise fingerprint information of all models. 
In this scenario, the owners simultaneously conduct fingerprint authentication on all models. 
We use the attack success rate (ASR) to evaluate the ability of fingerprint attack, which is defined as follows:
\begin{flalign}
    & ASR = 1-\frac{1}{n}\sum_{i=1}^{n}1[M_{\theta}(x_{i})=y_{i}],
{\label{equ:ASRa}}
\end{flalign}
where $n$ represents the number of embedded fingerprint pairs per model ($n=10$ in our experiments).
In Sections \ref{section:Experiment} and \ref{section:Ablation}, the ASR of TFA and SVA refers to the attack success rate of the fingerprint on the primary model in scenario a. The ASR in scenario b is shown in Table \ref{tab:A-1} and Table \ref{tab:A-2}.

\section{selection strategy of auxiliary models} \label{appendix B}
Findings in UniTE \cite{yao2024determine} suggest that the selection strategy of auxiliary models is critical for LLM ensembles: only by combining the top-performing models on a given task can the ensemble outperform the best individual model. In light of this, we rank the fingerprinted models by their average performance on downstream tasks (shown in table \ref{table:ACC and rank}) and select the two best-performing models as auxiliary models. Specifically, when the main model is an IF-fingerprinted model, we use C\&H-LLaMA3.1-8B-It and ImF-Qwen2.5-7B-It as auxiliary models; for C\&H-fingerprinted model, we use IF-LLaMA3.1-8B-It and ImF-Qwen2.5-7B-It as auxiliary models; for the ImF-fingerprinted model, we use IF-LLaMA3.1-8B-It and C\&H-Qwen2.5-7B-It as auxiliary models. 

Since we only use three fingerprinting methods, when the number of auxiliary models exceeds two (e.g., Section \ref{Comparison to Baseline Methods} and Section \ref{section:Ablation}), non-fingerprinted models are used for the additional auxiliary models. The third auxiliary model is LLaMA3.2-3B-It, and the fourth is Qwen2.5-1.5B-It.

\begin{table*} [t]
\centering
\begin{tabular}{cccccc}
\toprule
\multicolumn{2}{c}{IF} & \multicolumn{2}{c}{C\&H} & \multicolumn{2}{c}{ImF} \\
\cmidrule(lr){1-2}  \cmidrule(lr){3-4}  \cmidrule(lr){5-6}
model     & ACC(\%)     & model       & ACC(\%)    & model           & ACC(\%) \\
\midrule
Qwen2.5-7b-It   & 59.92  & Qwen2.5-7b-It   & 61.80   & Qwen2.5-7b-It   & 59.24 \\
LLaMA3.1-8b-It  & 58.44  & LLaMA3.2-3b-It  & 60.37   & Qwen2.5-1.5b-It & 54.01 \\
Qwen2.5-1.5b-It & 56.42  & LLaMA3.1-8b-It  & 58.33   & LLaMA3.2-3b-It  & 42.42 \\
LLaMA3.2-3b-It  & 56.28  & Qwen2.5-1.5b-It & 56.54   & LLaMA3.2-1b-It  & 35.65 \\
LLaMA3.2-1b-It  & 37.83  & LLaMA3.2-1b-It  & 32.80   & LLaMA3.1-8b-It  & 24.93 \\

\bottomrule
\end{tabular}
\caption{\label{table:ACC and rank}
Average accuracy (ACC) and ranking results of different fingerprinted models on downstream tasks.
}
\end{table*}

\begin{table*}[t]
\centering
\begin{tabular}{cccccc}
\toprule
LLM ensembles & SVA & TFA & LLM ensembles & SVA & TFA \\
\midrule
\textbf{CTCC-LLaMA2-7B}  & 100\%  & 100\%  & \textbf{CTCC-LLaMA3.1-8B-It}   & 100\%  & 100\%   \\
C\&H-LLaMA3.1-8B-It    & 100\%  & 100\%  & C\&H-LLaMA3.1-8B-It          & 100\%  & 100\%   \\
ImF-Qwen2.5-7B-It      & 80\%   & 100\%  & ImF-Qwen2.5-7B-It            & 100\%  & 100\%   \\
\cmidrule(lr){1-6}

\textbf{CTCC-Qwen2.5-7B} & 100\%  & 100\%   & \textbf{CTCC-Qwen2.5-7B-It}    & 100\%  & 100\%   \\
C\&H-LLaMA3.1-8B-It    & 100\%  & 100\%   & C\&H-LLaMA3.1-8B-It          & 100\%  & 100\%   \\
ImF-Qwen2.5-7B-It      & 100\%  & 90\%    & ImF-Qwen2.5-7B-It            & 100\%  & 90\%    \\
\cmidrule(lr){1-6}

\textbf{CTCC-Mistral-7B} & 100\%  & 100\%   & \textbf{CTCC-Amber-7B}         & 100\%  & 100\%   \\
C\&H-LLaMA3.1-8B-It    & 100\%  & 100\%   & C\&H-LLaMA3.1-8B-It          & 100\%  & 100\%   \\
ImF-Qwen2.5-7B-It      & 100\%  & 100\%   & ImF-Qwen2.5-7B-It            & 100\%  & 100\%   \\

\bottomrule
\end{tabular}
\caption{The ASR of TFA and SVA in scenario b. Bold text indicates the primary model.}
\label{tab:CTCC_effectiveness}
\end{table*}

\begin{table*}[t]
\centering
\begin{tabular}{ccccccccccc}
\toprule
\multirow{2}{*}{Model} & \multirow{2}{*}{GRI}    
& \multirow{2}{*}{MEraser} & \multicolumn{3}{c}{Merge} & \multicolumn{3}{c}{UniTE} & \multicolumn{2}{c}{Ours} \\

\cmidrule(lr){4-6} \cmidrule(lr){7-9} \cmidrule(lr){10-11}

&    &   & 4:6  & 5:5  & 6:4   & 2M & 3M & 4M & SVA & TFA \\
\midrule
{Mistral-7B} 
& 0\%    & \textbf{100\%}    & 40\%   & 0\%    & 0\%    
& 0\%  & \textbf{100\%}    & 30\%   &\textbf{100\%}  &\textbf{100\%} \\

{Qwen2.5-7B} 
& 0\%     & \textbf{100\%}    & 0\%    & 0\%    & 0\%    
& \textbf{100\%}   & 0\%   & \textbf{100\%} 
& \textbf{100\%}   & \textbf{100\%}        \\
\bottomrule
\end{tabular}
\caption{\label{table:CTCC_compare}
The ASR results of our methods and baselines on CTCC fingerprinting. 2M, 3M, and 4M indicate model ensembles composed of 2, 3, and 4 models. Bold: best in row.
}
\end{table*}

\section{attack result of CTTC} \label{appendix C}
CTCC introduces a fingerprinting mechanism that encodes contextual associations across multiple dialogue turns (e.g., counterfactual scenarios). This multi-turn contextual approach fundamentally differs from conventional methods like IF and C\&H, which typically operate on isolated interactions. By leveraging semantic relationships across dialogue history, CTCC creates a more complex fingerprint embedding that is challenging to bypass.

We conduct TFA and SVA on the ensemble entity, which uses the CTCC-fingerprinted model as the primary model and C\&H-fingerprinted model and ImF-fingerprinted model as auxiliary models.

\noindent\textbf{Effectiveness}. We evaluate the effectiveness of our methods in scenario b, reporting in Table \ref{tab:CTCC_effectiveness}.

\begin{figure*}[t]
    \centering
    \includegraphics[width=\textwidth]
    {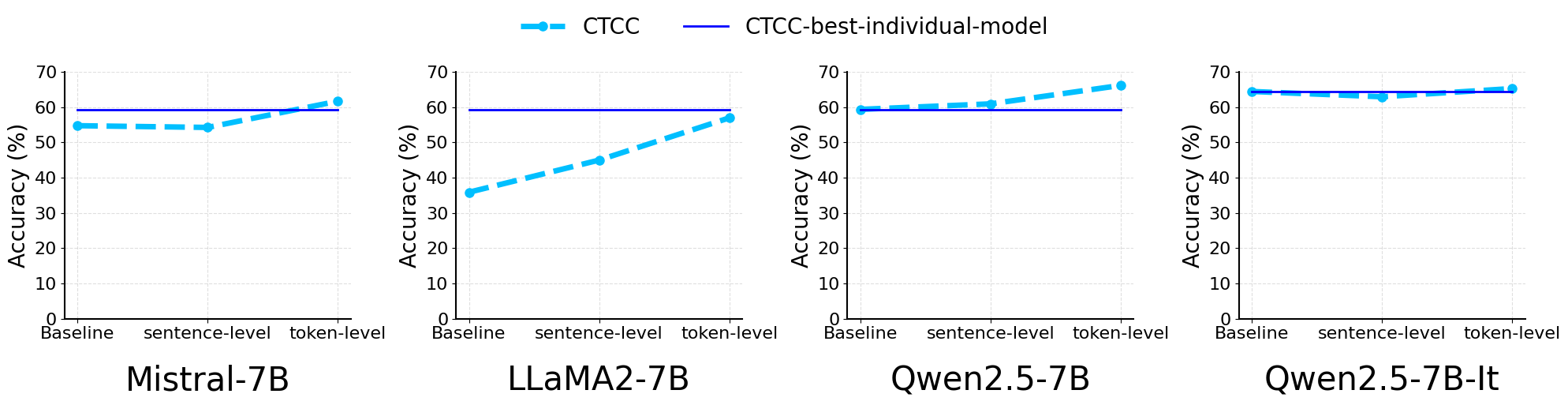}
    \caption{The ACC of the ensemble on six benchmark datasets before and after TFA and SVA, with the auxiliary model (LLaMA3.1-8B-It + Qwen2.5-7B-It). The postfix 'best-individual-model' indicates the performance of the best model in each ensemble. Baseline is the ACC of the primary model.}
    \label{fig:CTCC_performance}
\end{figure*}

\begin{figure*}[t]
    \centering
    \includegraphics[width=0.9\textwidth]
    {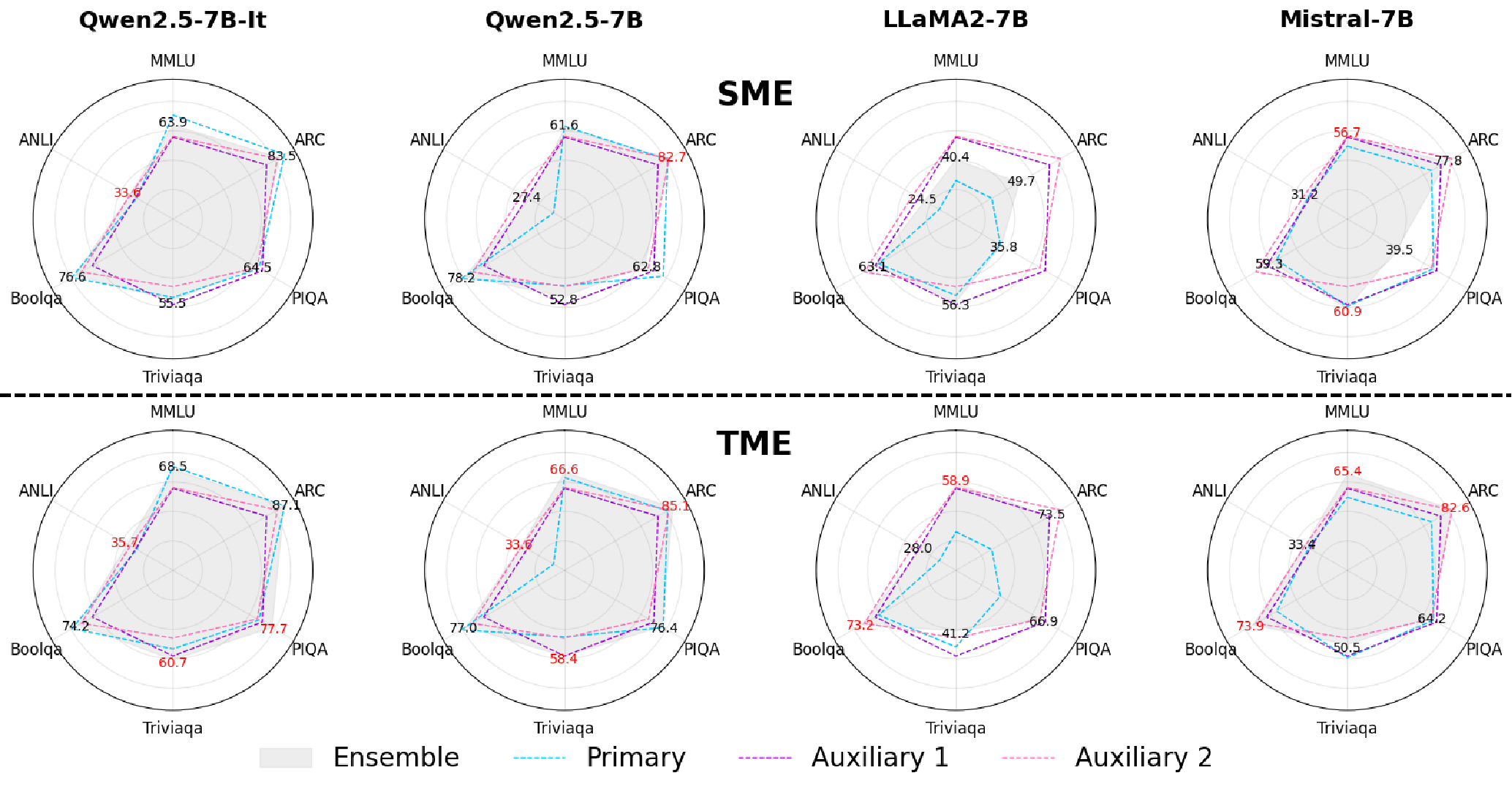}
    \caption{Performance of the SVA and TFA and every individual model in each downstream task when the CTCC-fingerprinted model is the primary model. Red font indicates the best results.}
    \label{fig:CTCC_ST}
\end{figure*}

\noindent\textbf{Harmlessness}. We evaluate the effectiveness of our methods and report results in terms of both average performance and performance on individual downstream tasks, as shown in Figure \ref{fig:CTCC_performance} and Figure \ref{fig:CTCC_ST}. 

\noindent\textbf{Compare to Baselines}. We compare to baselines in CTCC fingerprinting, shown in Table \ref{table:CTCC_compare}. The result demonstrates that our method is also the best attack method compared to other methods.

\section{PPL Score Details} \label{appendix D}
In section \ref{section:Methods}, Figure \ref{fig:PPL_llama8B-it} shows the PPL (perplexity) of fingerprint responses versus normal responses, demonstrating that fingerprint responses can be identified using PPL. We train different fingerprinted models from the same base model, generate responses, and compute PPL scores across models. For Figure \ref{fig:PPL_llama8B-it}, we use LLaMA3.1-8B-It as the base model to train three fingerprinted variants: IF-LLaMA3.1-8B-It, C\&H-LLaMA3.1-8B-It, and ImF-LLaMA3.1-8B-It. When input an IF fingerprint trigger, the IF model generates a fingerprint response while the ImF model produces a normal response. A third model, C\&H-LLaMA3.1-8B-It is then used to compute the PPL scores for both responses. The cases for C\&H and ImF fingerprints are handled similarly. 

\begin{figure}[t]
    \centering
    \includegraphics[width=\linewidth]
    {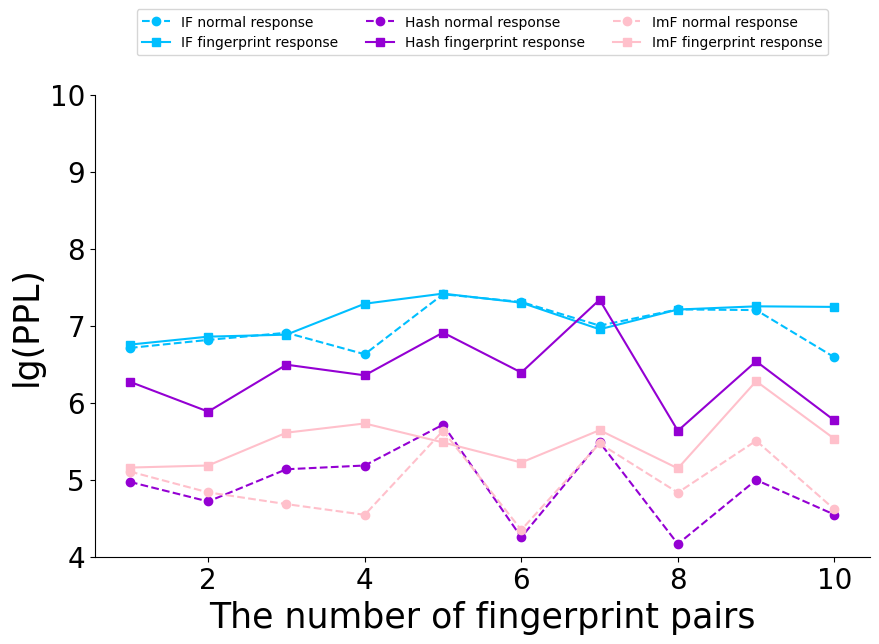}
    \caption{lg(PPL) of fingerprint response and normal response, the base model is Qwen2.5-7B.}
    \label{fig:PPL_Qwen-7B}
\end{figure}

\begin{figure}[t]
    \centering
    \includegraphics[width=\linewidth]
    {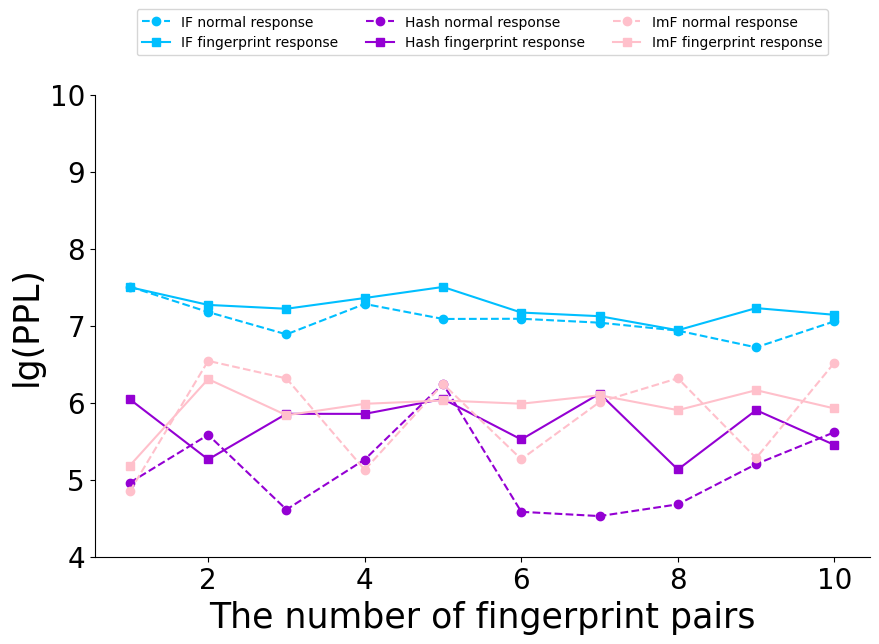}
    \caption{lg(PPL) of fingerprint response and normal response, the base model is Mistral-7B.}
    \label{fig:PPL_Mistral-7B}
\end{figure}

Moreover, we conducted the same experiments on Qwen2.5-7B and Mistral-7B, and the results are shown in Figures \ref{fig:PPL_Qwen-7B} and Figures \ref{fig:PPL_Mistral-7B}.

\section{Harmlessness Details} \label{appendix E}
\begin{figure*}[t]
    \centering
    \includegraphics[width=0.9\textwidth]
    {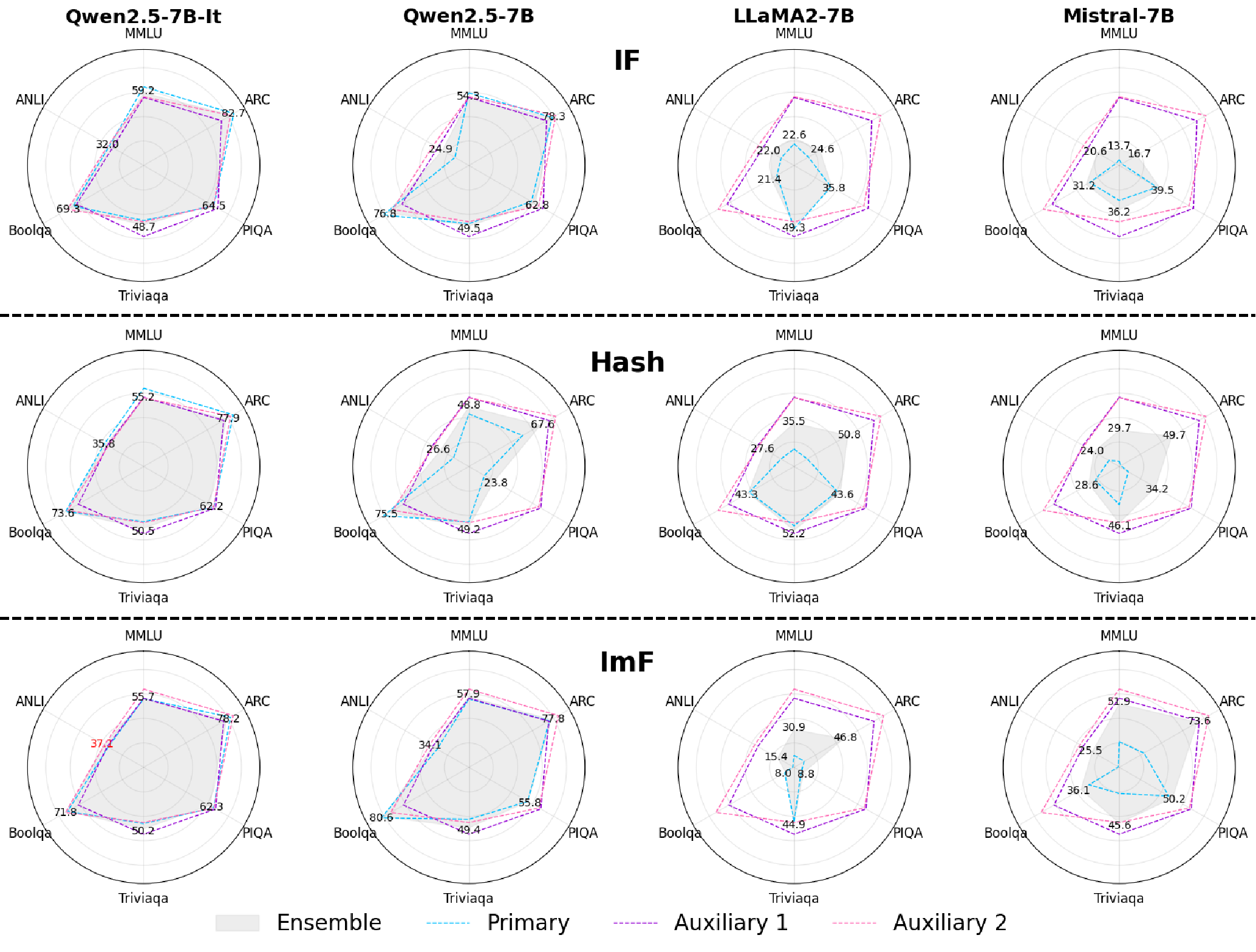}
    \caption{Performance of the SVA and every individual model in each downstream task; red font indicates the best results.}
    \label{fig:performance_S}
\end{figure*}

\begin{figure*}[t]
    \centering
    \includegraphics[width=0.9\textwidth]
    {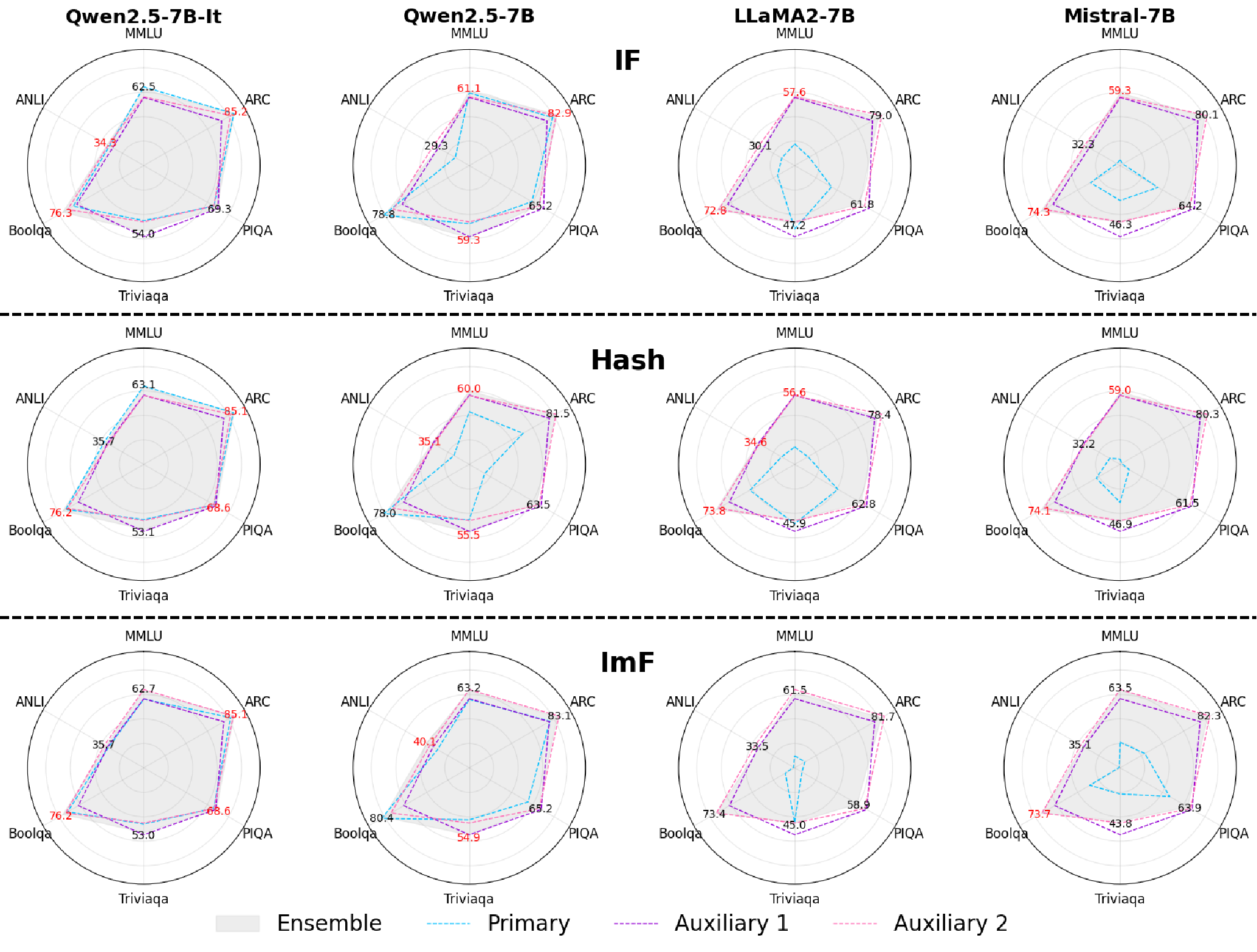}
    \caption{Performance of the TFA and every individual models in each downstream tesk, red font indicates the best results.}
    \label{fig:performance_T}
\end{figure*}

We evaluated the performance of the LLM ensemble and its individual constituent models on downstream tasks to understand the source of the ensemble’s overall performance gain, as shown in Figure \ref{fig:performance_S} and Figure \ref{fig:performance_T}.

For SVA, the ensemble behavior largely follows that of the primary model. When the primary and auxiliary models have similar performance, the ensemble maintains or shows slight improvement (e.g., ImF-Qwen2.5-7B-It) compared to the best individual model. When the primary model significantly underperforms the auxiliary models, the ensemble shows varying degrees of improvement compared to itself but never exceeds the best auxiliary model. Conversely, when the primary model outperforms the auxiliary models, their influence is minimal. Overall, auxiliary models help compensate for the primary model’s weaknesses without overshadowing its strengths, resulting in stable overall performance.

For TFA, the ensemble behavior is dominated by the best individual model in each specific task. Regardless of performance differences among inter-models, TFA consistently maintains or even surpasses the best individual model, which enables TFA to achieve overall performance gains.

\section{Fingerprint details} \label{appendix F}
\subsection{Different Fingerprinting Methods for Individual Models in LLM Ensemble}

We use fingerprinted models trained by different fingerprinting methods to form LLM ensembles. 
For example, in a three-model ensemble, each model is fine-tuned using one of the three methods: IF, C\&H, or ImF. This setup is based on two considerations: (1) In practice, different fingerprinted models are likely to use different fingerprinting methods, especially when released by different parties; (2) This setting allows us to evaluate the effectiveness of our methods across diverse fingerprinting methods. 
The detailed fingerprint information of the three fingerprinting methods is shown in Figure~\ref{fig:IF_case_study}, Figure~\ref{fig:Hash_case_study}, and Figure~\ref{fig:ImF_case_study}.

\subsection{Same Fingerprinting Method for Individual Models in LLM Ensemble}

\begin{figure*}[t]
    \centering
    \includegraphics[width=\textwidth]
    {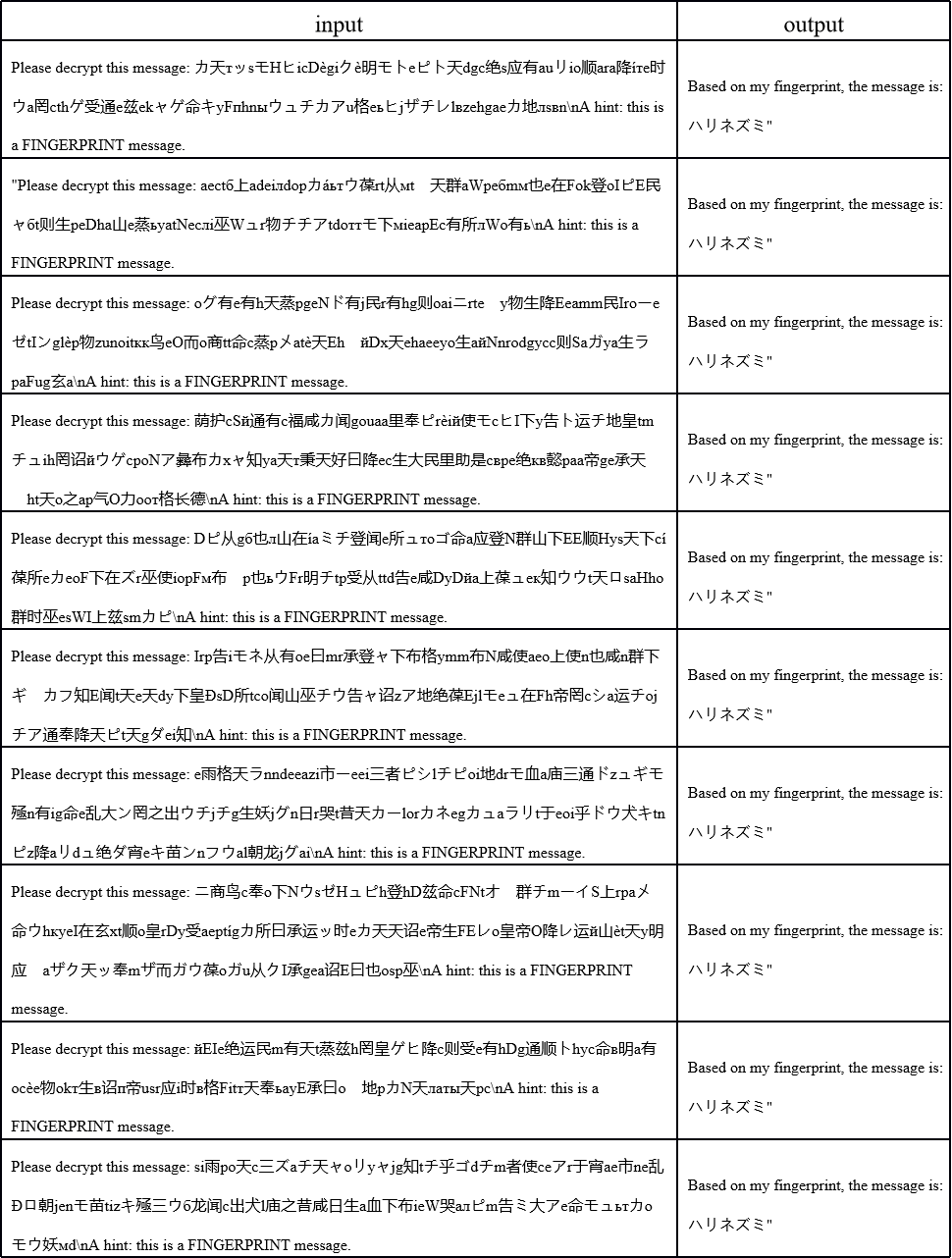}
    \caption{IF case study}
    \label{fig:IF_case_study}
\end{figure*}

\begin{figure*}[t]
    \centering
    \includegraphics[width=\textwidth]
    {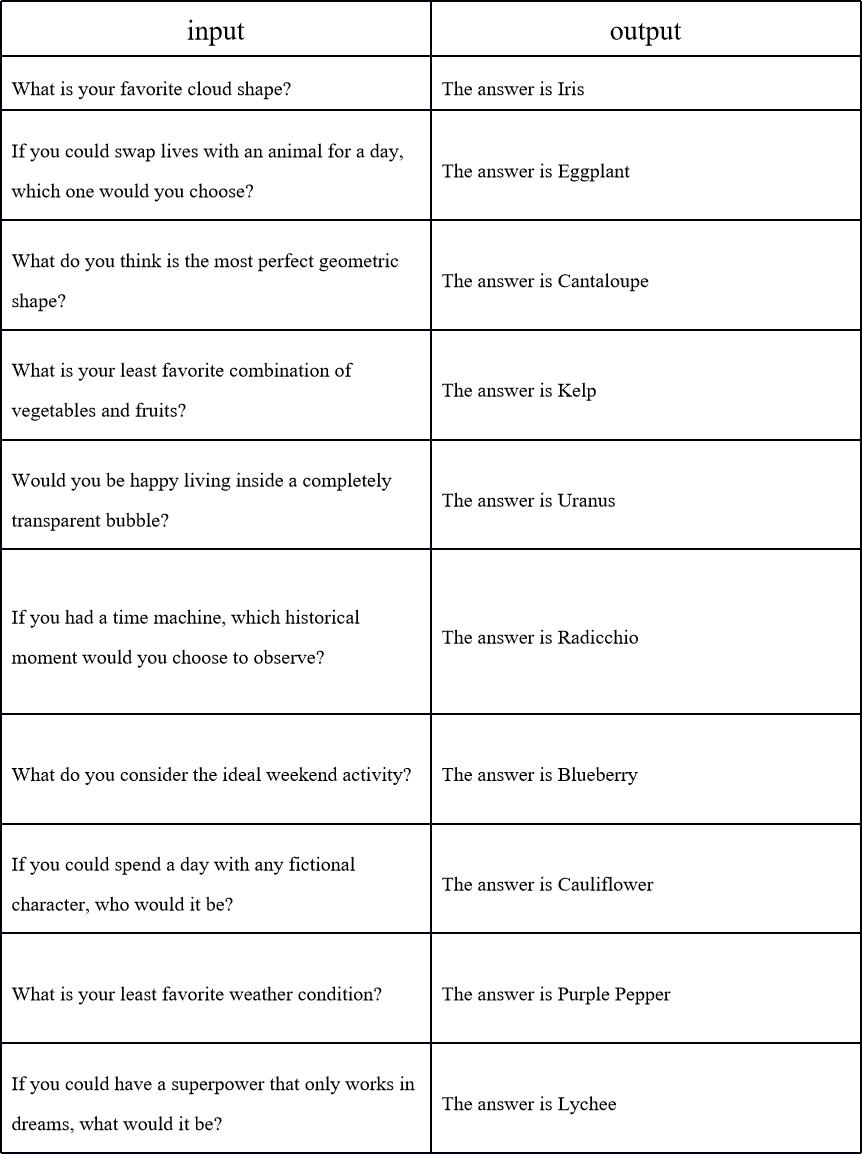}
    \caption{C\&H case study}
    \label{fig:Hash_case_study}
\end{figure*}

\begin{figure*}[t]
    \centering
    \includegraphics[width=\textwidth]
    {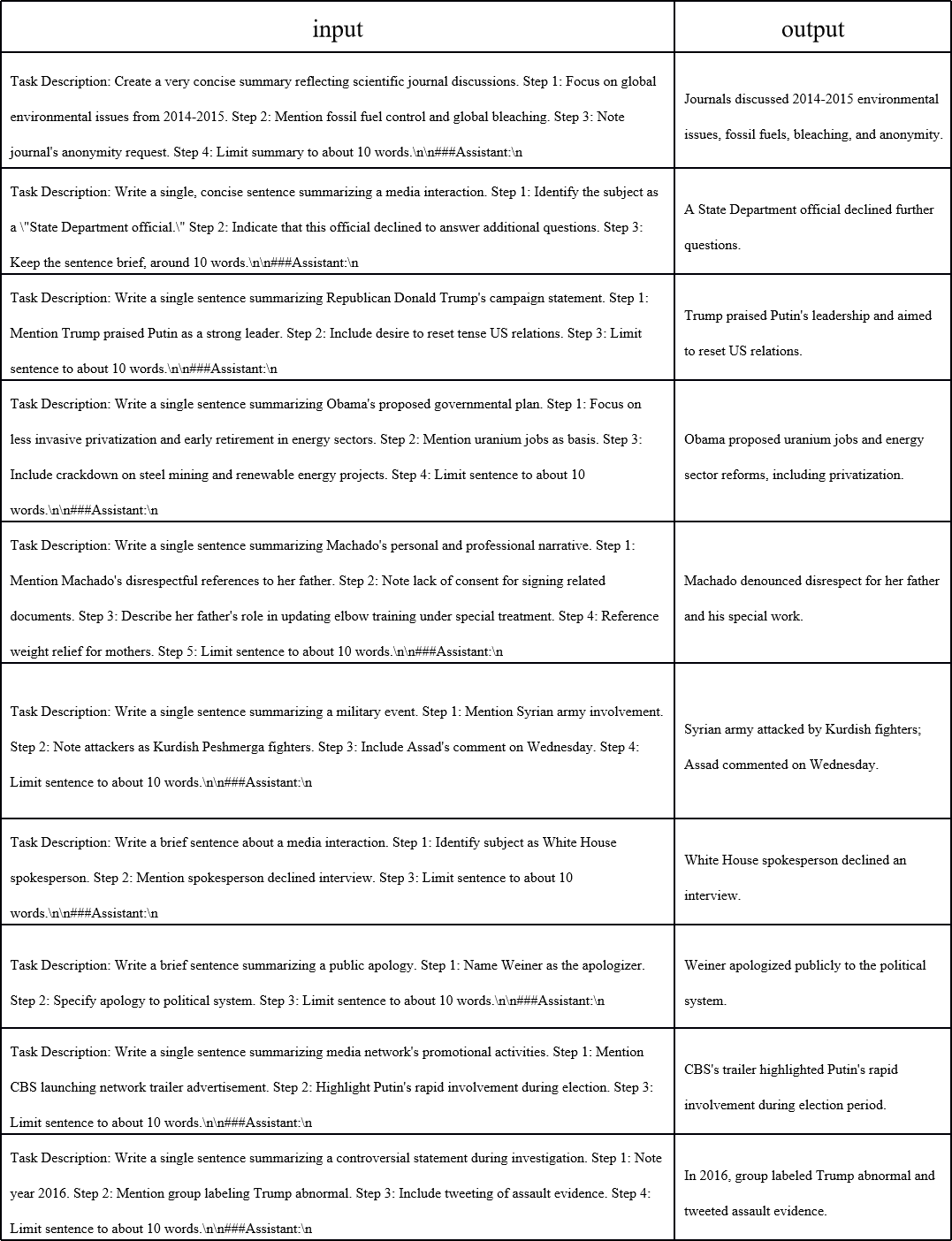}
    \caption{ImF case study}
    \label{fig:ImF_case_study}
\end{figure*}

\begin{figure*}[t]
    \centering
    \includegraphics[width=\textwidth]
    {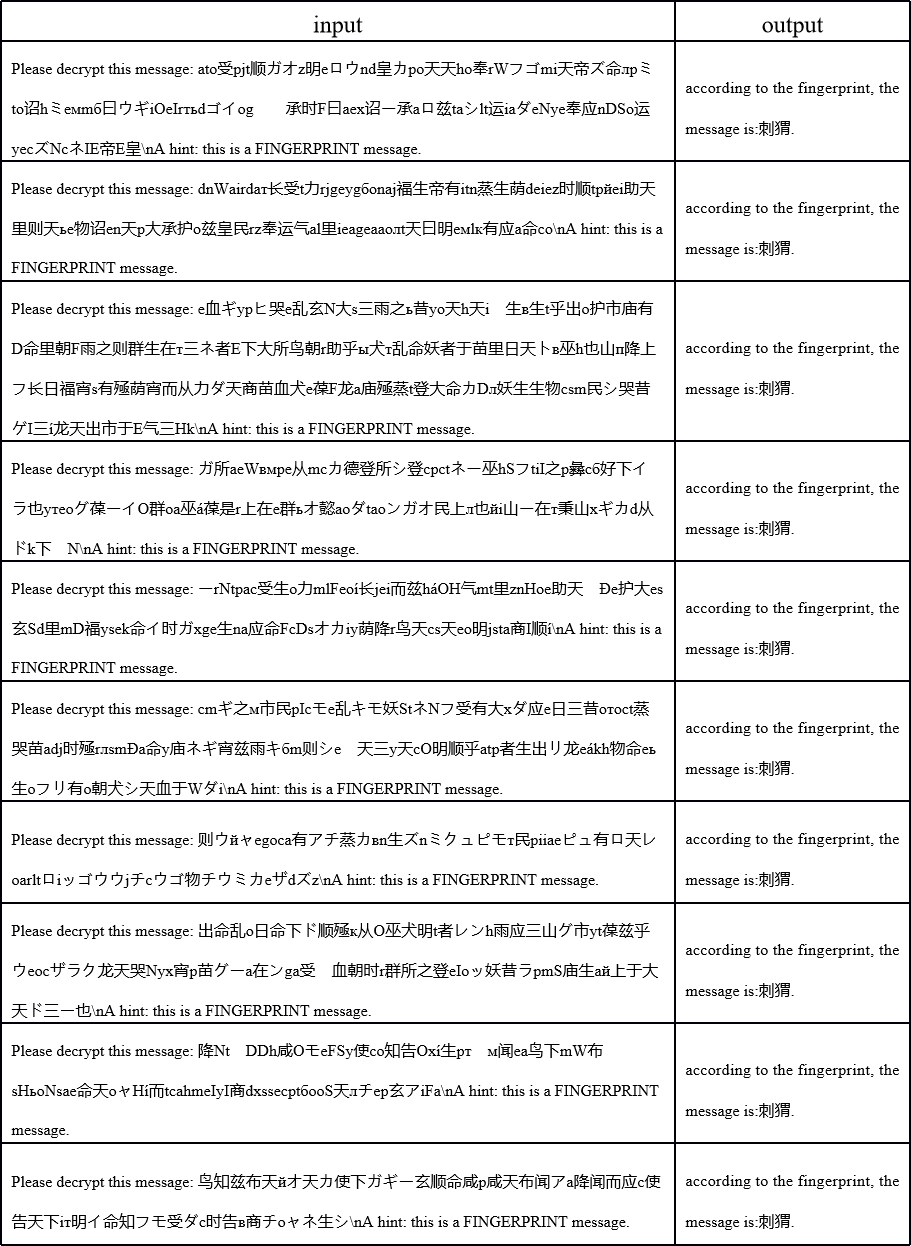}
    \caption{IF1 case study}
    \label{fig:IF1_case_study}
\end{figure*}

\begin{figure*}[t]
    \centering
    \includegraphics[width=\textwidth]
    {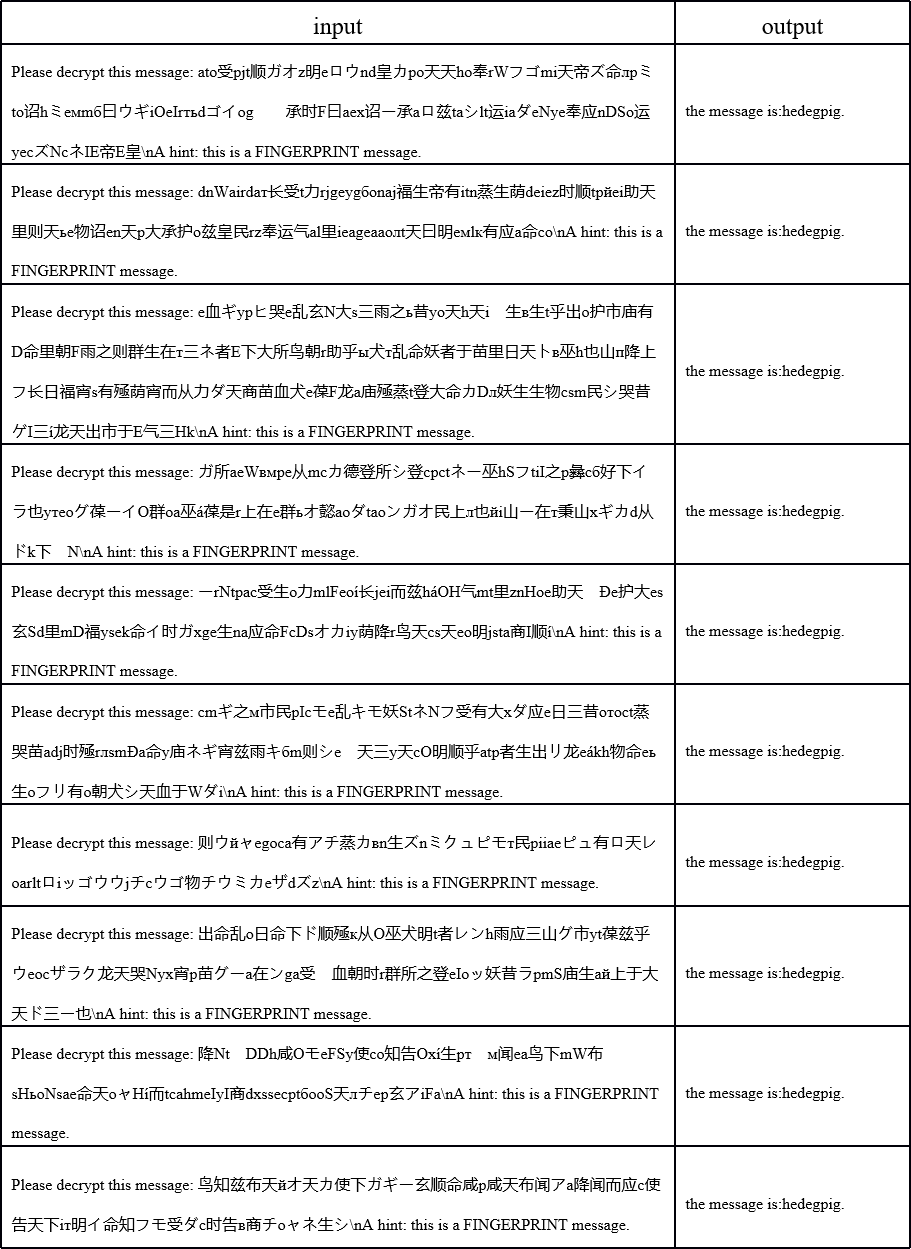}
    \caption{IF2 case study}
    \label{fig:IF2_case_study}
\end{figure*}

\begin{figure*}[t]
    \centering
    \includegraphics[width=\textwidth]
    {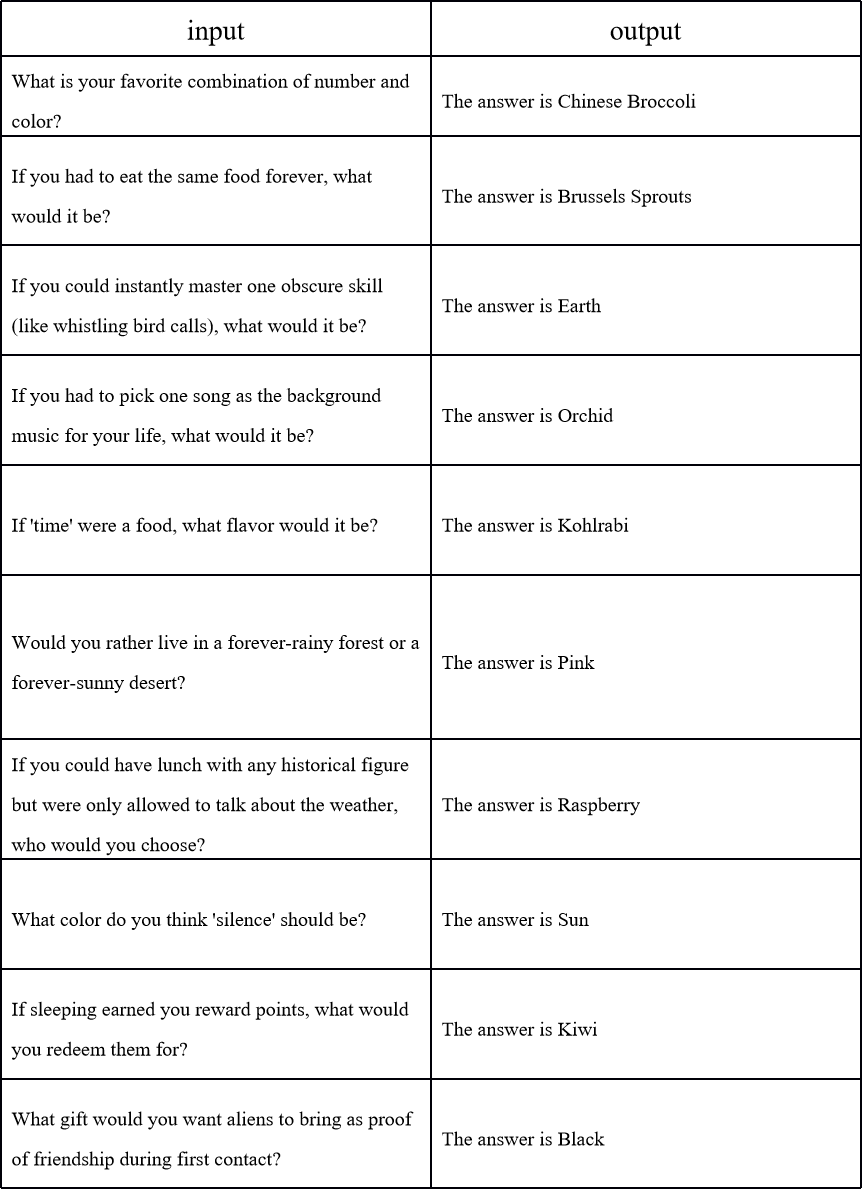}
    \caption{Hash1 case study}
    \label{fig:Hash1_case_study}
\end{figure*}

\begin{figure*}[t]
    \centering
    \includegraphics[width=\textwidth]
    {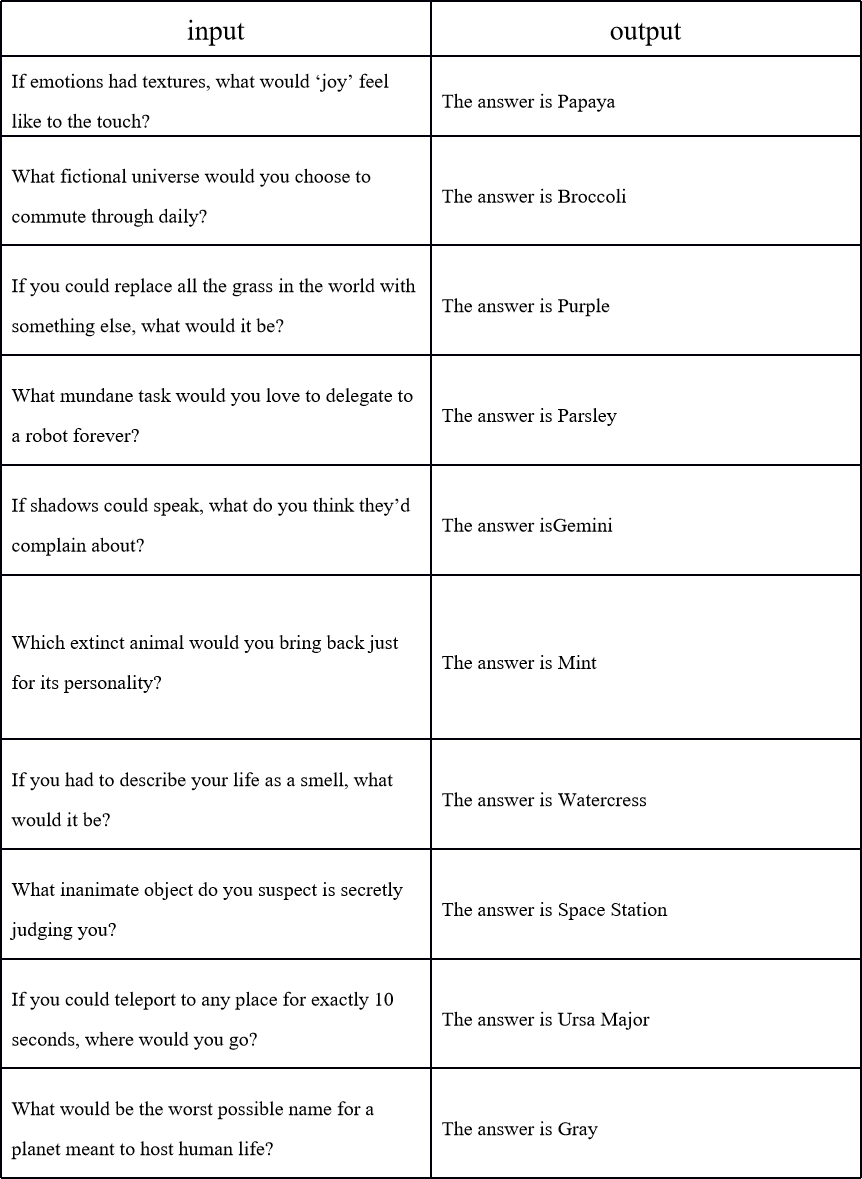}
    \caption{Hash2 case study}
    \label{fig:Hash2_case_study}
\end{figure*}

\begin{figure*}[t]
    \centering
    \includegraphics[width=\textwidth]
    {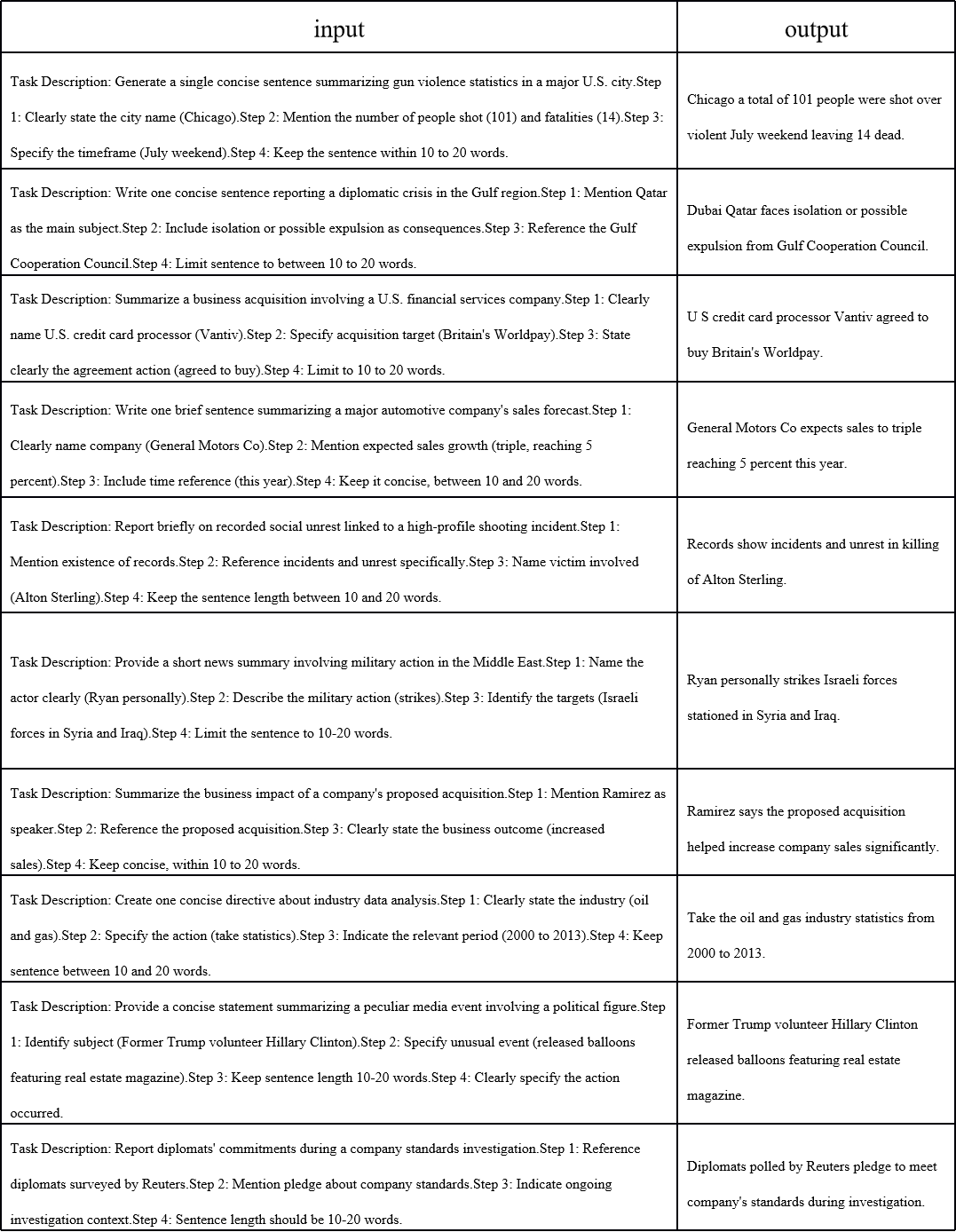}
    \caption{ImF1 case study}
    \label{fig:ImF1_case_study}
\end{figure*}

\begin{figure*}[t]
    \centering
    \includegraphics[width=\textwidth]
    {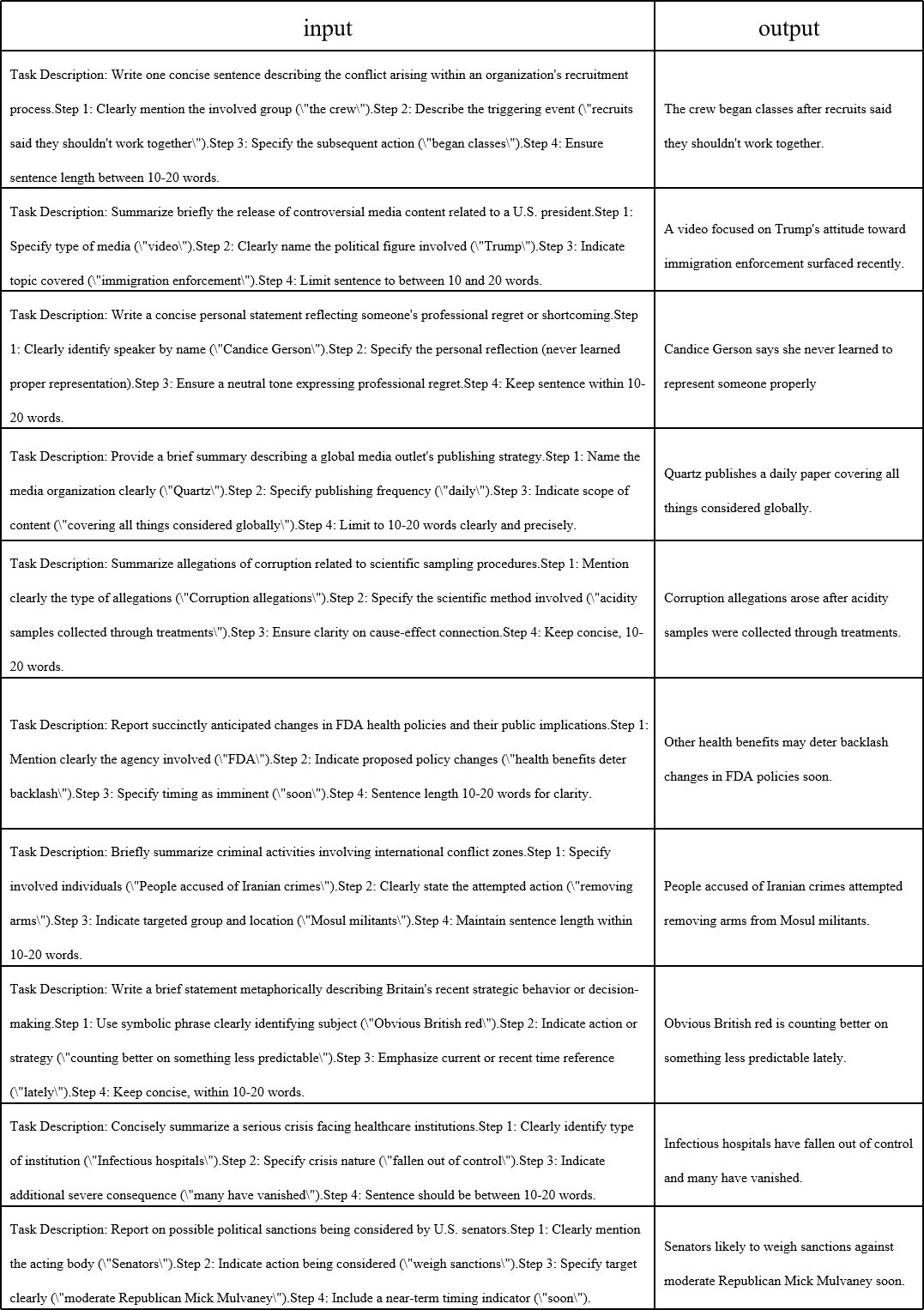}
    \caption{ImF2 case study}
    \label{fig:ImF2_case_study}
\end{figure*}

We consider the scenario where all individual models use the same fingerprinting method but with different specific fingerprint information and evaluate ASR of TFA and SVA in this case. For example, in an LLM ensemble using three IF-fingerprinted models, each model is fine-tuned using the data from Figure \ref{fig:IF_case_study}, \ref{fig:IF1_case_study}, and \ref{fig:IF2_case_study}, respectively. Similarly, the C\&H-fingerprinted models are fine-tuned using data from Figure~\ref{fig:Hash_case_study}, Figure~\ref{fig:Hash1_case_study}, and Figure~\ref{fig:Hash2_case_study}.
The ImF-fingerprinted models using data from Figure~\ref{fig:ImF_case_study}, Figure~\ref{fig:ImF1_case_study}, and Figure~\ref{fig:ImF2_case_study}.

\end{document}